\documentclass[useAMS,usenatbib]{mn2e}
\usepackage{graphicx}
\usepackage{amsmath}
\usepackage{hyperref}
\usepackage{subfigure}
\usepackage{txfonts}

\usepackage{color}
\definecolor{darkgreen}{rgb}{0,0.6,0}
\definecolor{gray}{rgb}{0.5,0.5,0.5}
\definecolor{mauve}{rgb}{0.58,0,0.82}

\usepackage{hyperref}
\hypersetup{breaklinks,colorlinks,citecolor=blue}

\newcommand{\compressivesensing}{compressive sensing}
\newcommand{\Compressivesensing}{Compressive sensing}
\newcommand{\pc}{\ percent}

\newif\ifdraft
\drafttrue

\ifdraft
  \newcommand{\before}[1]{\textcolor{darkgreen}{[{\bf BEFORE}: #1]}}

\else
  \newcommand{\before}[1]{}
  
\fi

\newcommand{\eg}{\mbox{\it e.g.}}
\newcommand{\ie}{\mbox{\it i.e.}}

\title[Revisiting the spread spectrum effect]{Revisiting the spread spectrum effect in radio interferometric imaging: a sparse variant of the $w$-projection algorithm}
\author[L. Wolz \emph{et al.}]{L. Wolz$^{1,}$$^2$\thanks{lwolz@star.ucl.ac.uk}, J. D. McEwen$^{1,}$$^3$, F. B. Abdalla$^1$, R. E. Carrillo$^4$, Y. Wiaux$^{4,}$$^{5,}$$^{6,}$$^7$\\
$^1$Department of Physics and Astronomy, University College London, London WC1E 6BT, UK\\
$^2$ Sub-Department of Astrophysics, Department of Physics, University of Oxford,
The Denys Wilkinson Building,
Keble Road,
Oxford
OX1 3RH,
UK\\
$^3$ Mullard Space Science Laboratory, University College London, Surrey RH5 6NT, UK\\
$^4$ Institute of Electrical Engineering, 
Ecole Polytechnique Federale de Lausanne, 
CH-1015 Lausanne, 
Switzerland\\
$^5$ Department of Radiology and Medical Informatics, 
University of Geneva, CH-1211 Geneva, 
Switzerland\\
$^6$ Department of Radiology, 
Lausanne University Hospital, CH-1011 Lausanne, Switzerland\\
$^7$ Institute of Sensors, 
Signals \& Systems, Heriot Watt University, 
Edinburgh EH14 4AS, UK
}

\date{Accepted 2013 September 9. Received 2013 August 21; in original form 2013 July 12}
\pagerange{\pageref{firstpage}--\pageref{lastpage}} 
\pubyear{2013}

\begin{document}
\maketitle
\label{firstpage}
\begin{abstract}
Next-generation radio interferometric telescopes will exhibit non-coplanar baseline configurations and wide field-of-views, inducing a $w$-modulation of the sky image, which induces the spread spectrum effect. We revisit the impact of this effect on imaging quality and study a new algorithmic strategy to deal with the associated operator. In previous studies it has been shown that image recovery in the framework of compressed sensing is improved due to this effect, where the $w$-modulation can increase the incoherence between measurement and sparsifying signal representations. For the purpose of computational efficiency, idealised experiments with a constant baseline component $w$ were performed. We extend this analysis to the more realistic setting where the $w$-component varies for each visibility measurement. Firstly, incorporating varying $w$-components into imaging algorithms is a computational demanding task. We propose a variant of the $w$-projection algorithm, which is based on an adaptive sparsification procedure, and incorporate it in compressed sensing imaging methods. Secondly, we show that for varying $w$-components, reconstruction quality is significantly improved compared to no $w$-modulation, reaching levels comparable to a constant, maximal $w$-component. This finding confirms that one may seek to optimise future telescope configurations to promote large $w$-components, thus enhancing the fidelity of image reconstruction. 
\end{abstract}


\section{Introduction}
\label{sec-intro}


Future radio interferometric telescopes are planned to have increased sensitivity and wider field-of-views (FoV) than ongoing projects. Hence, imaging techniques must be able to handle more complex settings like wide FoVs and non-coplanar telescope alignments in order to ensure accurate image recovery. Assuming a small FoV and a coplanar baseline alignment, the imaging problem can be posed in two dimensions neglecting the spherical nature of the sky and Earth. For the next generation of radio interferometric surveys, these approximations are not reasonable and it is necessary to develop imaging techniques that model the radio interferometric measurement equation more accurately.

The generalisation from planar approximations to wide FoV settings has been considered previously \citep{cornwell:1992, Cornwell:2008qj, McEwen:2008ma, Carozzi:2008zd, Wiaux:2009eh, McEwen:2010cq, Cornwell:2012kv}. A common approach that is finding widespread application is the incorporation of the $w$-projection algorithm \citep{Cornwell:2008qj}. The $w$-projection algorithm reformulates the radio interferometric measurement equation in such a way that the two-dimensional Fourier transform remains applicable, thus leading to efficient computation; this approach has been incorporated effectively in the latest imagers \citep{Bhatnagar:2008fk, Tasse:2012vc}. In these imagers, a generalisation to the so-called $A$-projection algorithm is made to handle direction-dependent primary beams.

Recently, the radio interferometric imaging problem has been addressed in the context of compressive sensing \citep{Wiaux:2008ed, Wiaux:2009eh, Suksmono:2009vg, Wiaux:2009vr, wenger2010sparseri, McEwen:2010cq,  Li:2011wc, Li:2011wd, Carrillo:2012uv}, where wide FoVs have sometimes been considered \citep{Wiaux:2009eh,  McEwen:2010cq}. In particular, the spread spectrum effect that arises in the wide FoV setting was first studied by \cite{Wiaux:2009eh}, where it was shown both theoretically and empirically to enhance reconstruction quality in the \compressivesensing\ framework.  A non-negligible $w$-component (the component of the baseline in the pointing direction of the telescope) produces a $w$-modulation of the sky image, which acts to increase the incoherence between sensing and sparsity bases, thereby enhancing reconstruction quality.  In the first study of the spread spectrum effect performed by \cite{Wiaux:2009eh}, a constant $w$-component was examined in order to reduce  significantly the computational cost of image recovery.

In the current article we revisit the spread spectrum effect in a more realistic setting, where the $w$-component varies for each visibility measurement. We support varying $w$-components by incorporating a variant of the $w$-projection algorithm \citep{Cornwell:2008qj} into the compressed sensing radio interferometric imaging framework. We propose an adaptive sparsification procedure to recover a sparse matrix representation of the $w$-modulation kernel. By applying the $w$-modulation via a sparse matrix multiplication, the computational burden and memory requirements are reduced substantially even when no approximation is made.  The computational cost may be further reduced, for a small loss in reconstruction fidelity, by approximating the $w$-modulation kernel in a controlled manner. This procedure is generic and applicable for all types of direction dependent effects. We also extend the analysis to more realistic images and to different sparse dictionaries, while also considering different levels of visibility coverage.

The remainder of the article is organised as follows.  In Sec.~\ref{sec-ri}, we give a brief description of the radio interferometric measurement equation, recast the description in a discrete setting, and briefly review how to solve the imaging problem with compressed sensing methods. In Sec.~\ref{sec-w}, we describe a variant of the $w$-projection algorithm, incorporate it into the imaging framework and study its fast application. In Sec.~\ref{sec-sim}, we describe the experimental set-up of our simulations and present our results regarding the impact of the spread spectrum effect in the setting of varying $w$-components. In Sec.~\ref{sec-con}, we present the main conclusions of this work.

\section{Radio interferometric imaging}
\label{sec-ri}
In this section we describe the radio interferometric measurement equation in the most general form and then proceed to recast the equation in operator formalism. Solving the interferometric imaging problem by compressed sensing techniques is outlined in a general form in the final part of this section.

\subsection{Radio interferometric measurement equation}

A radio interferometer is comprised of an array of telescopes and measures the Fourier coefficients, the so-called visibilities $\mathcal{V}$, of an image on the sky, where the visibility coordinates are given by the relative position between the antenna pairs. The baseline components $(u,v,w)$ are measured in units of the wavelength $\lambda$ of the incoming signal. The components $u$ and $v$ specify the planar baseline coordinates, while the third component $w$ is associated with the basis vector of the coordinate frame pointing towards the direction of the center of the FoV of the telescope. 
The brightness of the sky $I$ can be described in the same coordinate frame as the baseline, with components $\bmath l= (l,m,n)$. Representing the celestial sphere by the unit sphere, the component $n$ can be expressed in terms of $(l,m)$ by $n(l,m)=\sqrt{1-l^2-m^2}$. The radio interferometric measurement equation for monochromatic, unpolarized signals reads
\begin{equation}
\mathcal{V}(u,v,w)=\int \frac{I(l,m)}{\sqrt{1-l^2-m^2}} \times e^{-2\pi i (ul+vm+w(\sqrt{1-l^2-m^2}-1))} \mathrm{d}l\mathrm{d}m
\label{equ-ri0}
\end{equation}
(without loss of generality we neglect the shape of the primary beam pattern; one may consider it essentially set to unity or absorbed into the sky image). See \cite{thompson04} for further detail. 

The $(n(l,m)-1)$ term in the complex exponential of Eq.~(\ref{equ-ri0}) can be neglected if the condition 
\begin{equation}
(l^2+m^2)w\ll 1
\label{equ-cond}
\end{equation}
is fulfilled \citep{thompson04}.  This is in general true for either coplanar baseline configurations which lead to $w=0$ (as realised by, \eg, the VLA\footnote{\url{http://www.vla.nrao.edu/}}; \citealt{Condon:1998iy}) or a small FoV such that $l$ and $m$ are  small. This approximation leads to the van Cittert-Zernike theorem (see \eg\ \citealt{thompson04}), which states that the visibilities measured by a radio interferometric telescope are related to the sky brightness by a two-dimensional Fourier transform under the small FoV approximation. 
Some of the next-generation radio interferometric telescopes (\eg\ the SKA\footnote{\url{http://www.skatelescope.org/}}; \citealt{Carilli:2004nx}) will have considerably larger field of views and longer baselines than current telescopes, such that the condition specified by Eq.~(\ref{equ-cond}) is no longer valid.

It is convenient to rewrite Eq.~(\ref{equ-ri0}) as
\begin{equation}
\mathcal{V}(u,v,w)= \int \frac{I(l,m) C(l,m;w)}{\sqrt{1-l^2-m^2}} e^{-2\pi i (ul+vm)}\mathrm{d}l\mathrm{d}m,
\label{equ-ri}
\end{equation}
where the $w$-modulation kernel is defined by 
\begin{equation}
C(l,m;w)\equiv e^{-2\pi i w(\sqrt{1-l^2-m^2}-1)}.
\label{equ-chirp}
\end{equation}
In the scope of this paper, we follow \cite{Wiaux:2009eh} and use the first order approximation of $C(l,m;w)$ given by the linear chirp
\begin{equation}
C_1(l,m;w)=e^{-\pi i w(l^2+m^2)},
\label{equ-chirp1}
\end{equation}
which is valid if \mbox{$(l^2+m^2)^2w\ll 1$}. Note that for constant $w$, Eq.~(\ref{equ-ri}) reduces to a two-dimensional Fourier transform since \mbox{$C(l,m;w)\equiv C(l,m)$}, leading to an efficient computational implementation.  The $w$-modulation kernel modulates the sky brightness image, giving rise to the spread spectrum effect \citep{Wiaux:2009eh}.  This was studied by \cite{Wiaux:2009eh} for a constant $w$-component, which reduces computational costs significantly.  To study the spread spectrum effect for varying $w$-component is computationally demanding; this is the issue addressed and studied in the current article. 

\subsection{Operator description}
\label{sec-ri-op}
We recast the radio interferometric measurement equation of Eq.~(\ref{equ-ri}) in discrete operator formalism (although the measurement equation can also be expressed in the Mueller matrix formalism; see, \eg, \citealt{Mueller1948,2011A&A...527A.106S,Tasse:2012vc}). We follow the same operator description as \cite{Wiaux:2009eh}, where a $w$-modulation with constant $w$-component is considered (this restriction will be relaxed in Sec.~\ref{sec-w}).
The visibilities $\mathcal{V}$ are assumed to be measured on a uniform grid in $(u,v)$, yielding the vector of measured visibilities $\bmath{y}$. Similarly, the sky brightness $I$ is discretised on a uniform grid, yielding the image pixels (concatenated into the vector) $\bmath x$. The radio interferometric measurement equation can then be written as
\begin{equation}
          \bmath{y}=\mathbfss M \mathbfss F \mathbfss C \mathbfss U\bmath x + \bmath n= \boldsymbol \Phi\bmath x +\bmath n,
           \label{equ-csri1}
 \end{equation}
where the masking operator $\mathbfss{M}$ specifies the visibility sampling pattern of the telescope, $\mathbfss{F}$ denotes the two-dimensional Fourier transform, which is implemented by a fast Fourier transform (FFT), $\mathbfss{C}$ denotes the diagonal $w$-modulation operator that implements modulation by Eq.~(\ref{equ-chirp1}) (for constant $w$) and $\mathbfss{U}$ denotes an upsampling operator. Multiplication of the image by the $w$-modulation term spreads the Fourier modes of the image. To avoid aliasing, we follow the approach taken by \cite{Wiaux:2009eh} and \cite{McEwen:2010cq} and introduce an upsampling operator $\mathbfss U$, which increases the supported band-limit by zero-padding the image in Fourier space. The noise $\bmath{n}$ models the instrumental noise of the telescope.  In this article, we do not consider visibility measurements at continuous coordinates, thus we neglect the gridding operator that would be required to model this setting; we leave a further investigation of the spread spectrum effect in the setting of continuous visibilities to future work. 

 \subsection{Solving the interferometric inverse problem}
\label{sec-ri-inverse}

Eq.~(\ref{equ-csri1}) can be solved for the image $\bmath x$ using techniques from \compressivesensing. The theory of \compressivesensing\ provides a flexible framework for sparse signal recovery (\citealt{Candes:2006ff, Candes:2006pi, Donoho:2006bs, Candes:2008kl}). \Compressivesensing\ techniques have been applied effectively to radio interferometric imaging in simulated settings (\eg\ \citealt{Wiaux:2008ed,Wiaux:2009eh,McEwen:2010cq,Carrillo:2012uv}), where reconstruction quality has been shown to be superior to conventional imaging algorithms.

\Compressivesensing\ techniques are applicable under the presumption that the signal $\bmath{x}$  has a sparse or compressible representation in a dictionary $\boldsymbol{\Psi}$, for instance wavelets. The columns of $\boldsymbol{\Psi}$ are given by the atoms making up the dictionary. The inverse problem formulated in Eq.~(\ref{equ-csri1}) can be solved by minimising the sparsity of the signal representation in the dictionary $\boldsymbol{\Psi}$, subject to a data fidelity constraint.  That is, by solving the optimisation problem
\begin{equation}
\min_{\bmath \alpha}\|\bmath{ \alpha}\|_1\text{ such that }\| \bmath y-\boldsymbol{\Phi\Psi} \bmath{\alpha}\|_2\leq \epsilon.
\label{equ-CO}
\end{equation}
Recall that the $\ell_1$ norm is defined as $\|\bmath \alpha\|_1 \equiv \sum_{i} |\alpha_i|$ and the $\ell_2$ norm as $\|\bmath \alpha\|_2 \equiv (\sum_{i} |\alpha_i|^2)^{1/2}$. The constraint $\epsilon$ may be related to a residual noise level estimator (see, \eg, \citealt{Wiaux:2008ed,Carrillo:2012uv}).  The optimisation problem specified by Eq.~(\ref{equ-CO}) can be solved using convex optimisation methods, such as Douglas-Rachford splitting \citep{2007ISTSP...1..564C}. 

In addition to the standard synthesis problem specified by Eq.~(\ref{equ-CO}), the alternative analysis framework and other variants may also be considered.  For example, average sparsity in the analysis setting, with an additional reweighting scheme, was considered by \cite{Carrillo:2012uv}, yielding the SARA algorithm, which was shown to provide excellent reconstruction fidelity for interferometric imaging.

For accurate reconstruction it is essential that the measurement basis, in case of radio interferometry essentially the Fourier basis, is incoherent with the sparsity dictionary.  The incoherence of the bases is proportional to the inverse of the maximum of the inner products of measurement and sparsity vectors (see, \eg, \citealt{Candes:2007fu}). Consequently, Dirac and Fourier bases are maximally incoherent.  For radio interferometry, the Dirac basis (\ie\ pixel basis) is thus essentially optimal in terms of coherence for signal reconstruction.  However, the Dirac basis may not necessary provide a sparse representation of the underlying image.  In practice, there is therefore often a trade-off between coherence and sparsity.

This trade-off can be mitigated by the spread spectrum effect \citep{Wiaux:2009eh}. Since the measurement basis can essentially be identified with the Fourier basis for radio interferometry, coherence is given by the maximum modulus of the Fourier coefficient of the sparsity atoms. The $w$-modulation corresponds to a norm-preserving convolution in Fourier space, spreading the spectrum of the sparsity atoms, thus reducing the maximum modulus of their Fourier coefficients and increasing incoherence.  The increased incoherence due to this spread spectrum effect acts to improve the fidelity of image reconstruction.  The universality of the spread spectrum effect has been demonstrated by \cite{Wiaux:2009eh}, such that optimal coherence can be reached (for Gaussian sources) in the limiting case of a large $w$-component.

 \section{Sparse variant of w-projection: theory and implementation}
 \label{sec-w}
In this section we describe how to support varying $w$-components in the radio interferometric imaging framework outlined in the preceeding section. Firstly, we describe the idea of the $w$-projection algorithm \citep{Cornwell:2008qj} and explain how it is incorporated in our imaging framework. Secondly, an adaptive sparsification procedure is presented to recover a sparse matrix approximation of the Fourier representation of the $w$-modulation kernel. This approach reduces the memory requirements and computation-time of solving the interferometric imaging problem considerably.

\subsection{$w$-projection}

The operator description of Eq.~(\ref{equ-csri1}) is only valid for a $w$-modulation with constant $w$, \ie\ a constant $w$-component. For a first examination of the spread spectrum effect this approach is convenient since it reduces the computational burden of a quantitative study substantially; this is the approach that was taken by \cite{Wiaux:2009eh}. In this work, we  simulate more realistic survey settings where we include varying $w$-components in the imaging process. To do this we apply a variant of the $w$-projection algorithm \citep{Cornwell:2008qj}.

By applying the Fourier convolution theorem to Eq.~(\ref{equ-csri1}), the $w$-modulation operator $\mathbfss{C}$ may be moved through the Fourier transform $\mathbfss{F}$, so that it is applied as a convolution in visibility space, expressed by the operator $\hat{\mathbfss{C}}$.  Each row of the matrix operator $\hat{\mathbfss{C}}$ is given by the (shifted) Fourier transform of the real space representation of the $w$-modulation $C(l,m;w)$. Naively, expressing the application of the $w$-modulation in this manner is computationally less efficient that the original formulation given by Eq.~(\ref{equ-csri1}).  However, applying the $w$-modulation in this manner via the $\hat{\mathbfss{C}}$ operator has two important advantages.  Firstly, a different $w$-component may be used to compute each row of $\hat{\mathbfss{C}}$; hence, a different $w$-component can be considered when computing the visibility corresponding to a particular $(u,v)$ component, while still exploiting the computational accessibility of a two-dimensional FFT to implement $\mathbfss{F}$.  Secondly, many of the elements of $\hat{\mathbfss{C}}$ will be close to zero; hence, a sparse matrix approximation of $\hat{\mathbfss{C}}$ can be used to speed up its computational application and reduce memory requirements significantly.

Following this formulation, the radio interferometric measurement equation to be solved may be written
 \begin{equation}
          \bmath{y}=\hat{ \mathbfss C} \mathbfss{F}\mathbfss{U}\bmath x + \bmath n= \boldsymbol \Phi^\prime\bmath{x} +\bmath n.
           \label{equ-csri2}
 \end{equation}
 where the other operators remain unchanged compared to Eq.~(\ref{equ-csri1}).
 The masking operator $\mathbfss{M}$ is incorporated in $\hat{\mathbfss{C}}$, where each row corresponds to one measured visibility. The resulting operator hence has dimension number of measured visibilities $M$ by number of pixels of the image $N^2$ (where we consider $N \times N$ pixel images).
We adopt this new description of the measurement operator $\boldsymbol \Phi^\prime$ when solving the radio interferometric optimisation problem. Note that the convex optimisation algorithms highlighted in Sec.~\ref{sec-ri-inverse} are transparent to the measurement operator $\boldsymbol \Phi$ and only require the application of the operator and its adjoint; hence, the $w$-projection algorithm may be incorporated into the compressed sensing framework easily.

In our description of the $w$-projection algorithm, every $(u,v)$ visibility component can have a different $w$-component. Contrast this to the approach taken typically, where, for computational reasons, the $w$-projection is performed on a relatively small number of $w$-planes.  We do not require such a restriction in our approach since we address computational issues by making a fast and accurate approximation of the application of the $\hat{\mathbfss{C}}$ operator, as described subsequently.

\subsection{Fast approximation of the $w$-modulation operator $\hat{\mathbfss{C}}$}

\begin{figure}
\includegraphics[width=0.5\textwidth, clip=true, trim= 50 0 0 0]{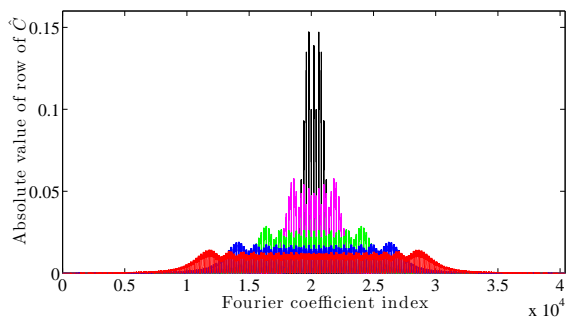}
\caption{Rows of the $w$-modulation kernel $\hat{\mathbfss{C}}$ for different values of \mbox{$w_{\rm d}=\{0.10,\ 0.25,\ 0.50,\ 0.75,\ 1.00\}$}, plotted in black, magenta, green, blue and red, respectively. The number of significant kernel entries and their support depends strongly on the level of $w$-modulation, \ie\ on the value of $w_{\rm d}$.  Consequently, the approximate kernel support should be chosen adaptively for each $w_{\rm d}$. }
\label{fig:chirpkernel}
\end{figure}

For every visibility component $(u,v,w)$, the $w$-modulation kernel is computed in real space via Eq.~(\ref{equ-chirp1}). Then the kernel is Fourier transformed via a two-dimensional FFT.  The resulting convolution kernel is collapsed into a vector which defines one (shifted) row of the $w$-modulation operator $\hat{\mathbfss{C}}$. 
In Fig.~\ref{fig:chirpkernel}, some exemplary rows of the absolute value of the complex convolution operator $\hat{\mathbfss{C}}$ are shown, where we quantify the amount of modulation by the variable $w_{\rm d}$, which is simply a rescaling of $w$. A value $w_{\rm d}=0$ implies no modulation, while $w_{\rm d}=1.0$ corresponds to maximal modulation (the explicit definition of  $w_{\rm d}$ is given in Sec.~\ref{subsec-setup}). As expected, it can be seen that the support and amplitude of the $w$-modulation kernel in Fourier space depends on $w_{\rm d}$ (since $w$ is related to its instantaneous frequency). Furthermore, the absolute value of a large number of the entries of rows of the convolution operator are very close to zero, particularly for small $w$-modulation. 

In order to speed up computational performance and reduce memory requirements, we sparsify  the $w$-modulation operator $\hat{\mathbfss{C}}$. Since the Fourier support of the $w$-modulation kernel varies for different $w_{\rm d}$, it is not efficient to \emph{a priori} limit the support of a row of the matrix operator to a constant size. Instead, the number of non-zero elements of each row of $\hat{\mathbfss{C}}$ should be adaptive to $w_{\rm d}$. 
Furthermore, we do not presume to know \emph{a priori} the location of the significant elements of $\hat{\mathbfss{C}}$; thus, the location of non-zero elements should also be chosen adaptively.
We therefore recover a sparse approximation of $\hat{\mathbfss{C}}$ by applying a different threshold for each row of $\hat{\mathbfss{C}}$: all entries less than the threshold (in absolute value) are set identically to zero. The threshold is chosen in such way that a certain fraction of the total energy of the $w$-modulation kernel is preserved.
The total energy is calculated by summing the squared elements of one row of the $w$-modulation kernel. 
The threshold of each matrix row is determined by considering the preserved energy distribution as a function of the threshold. We invert this relation in a computationally efficient way by using a bisecting method to determine the threshold for a specified level of preserved energy.
This is a very general approach to the sparsification of $\hat{\mathbfss{C}}$ since we do not make any assumption about the structure of each row; hence, our strategy is applicable for all types of direction-dependent effects and not only $w$-modulation.  

\begin{figure*}
\begin{tabular}{c c c c}

\includegraphics[width=.22\textwidth]{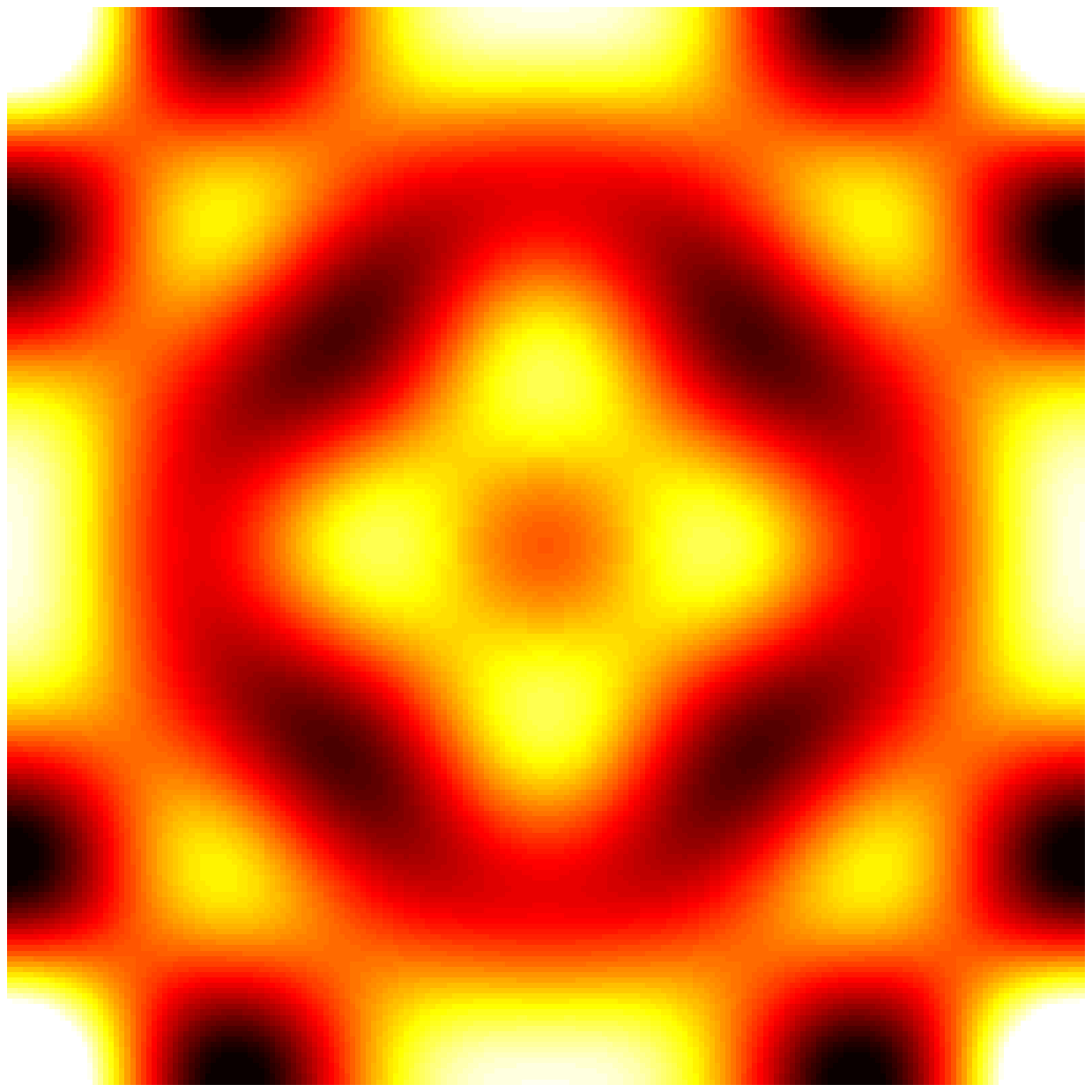} &
\includegraphics[width=.22\textwidth]{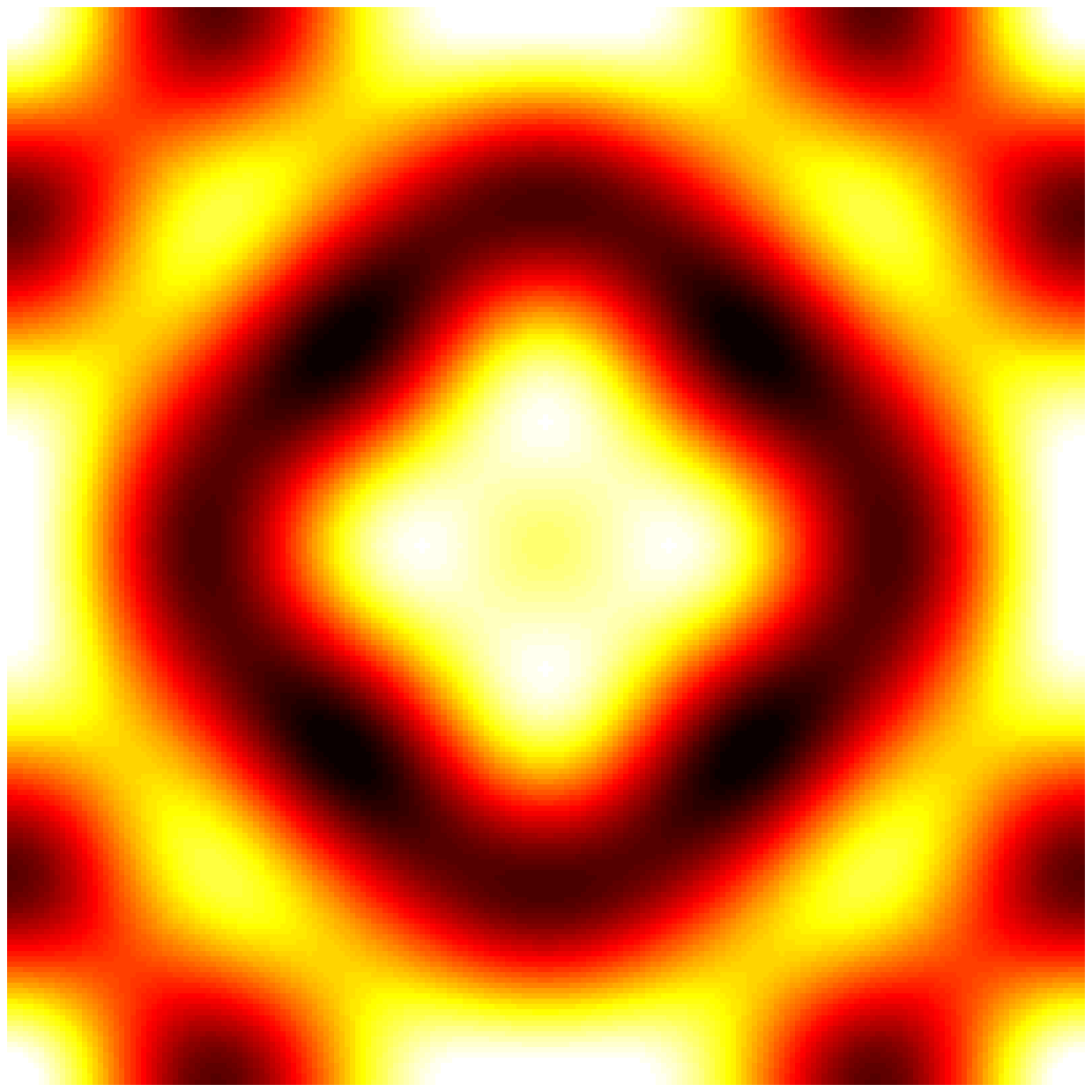} &
\includegraphics[width=.22\textwidth]{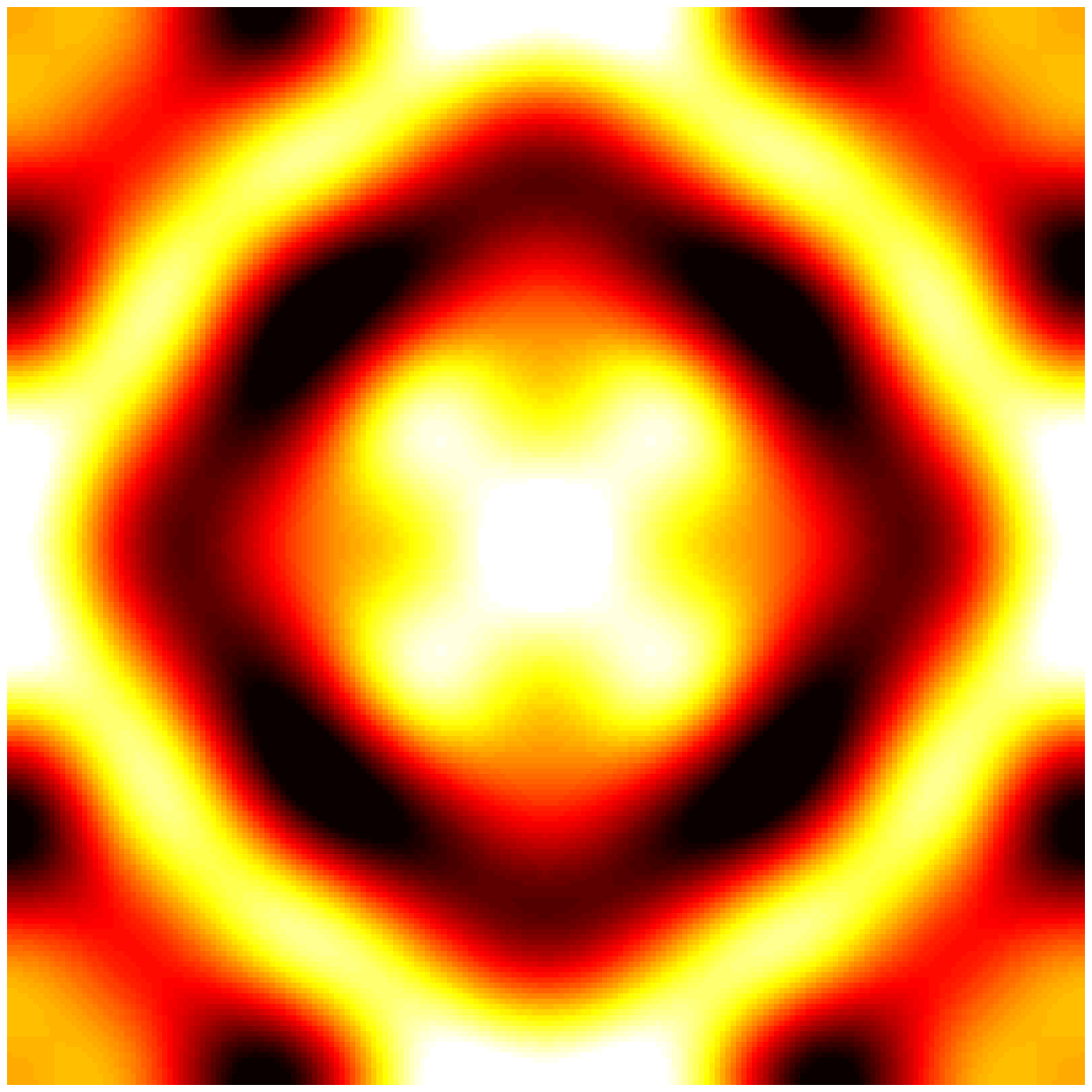} &
\includegraphics[width=.22\textwidth]{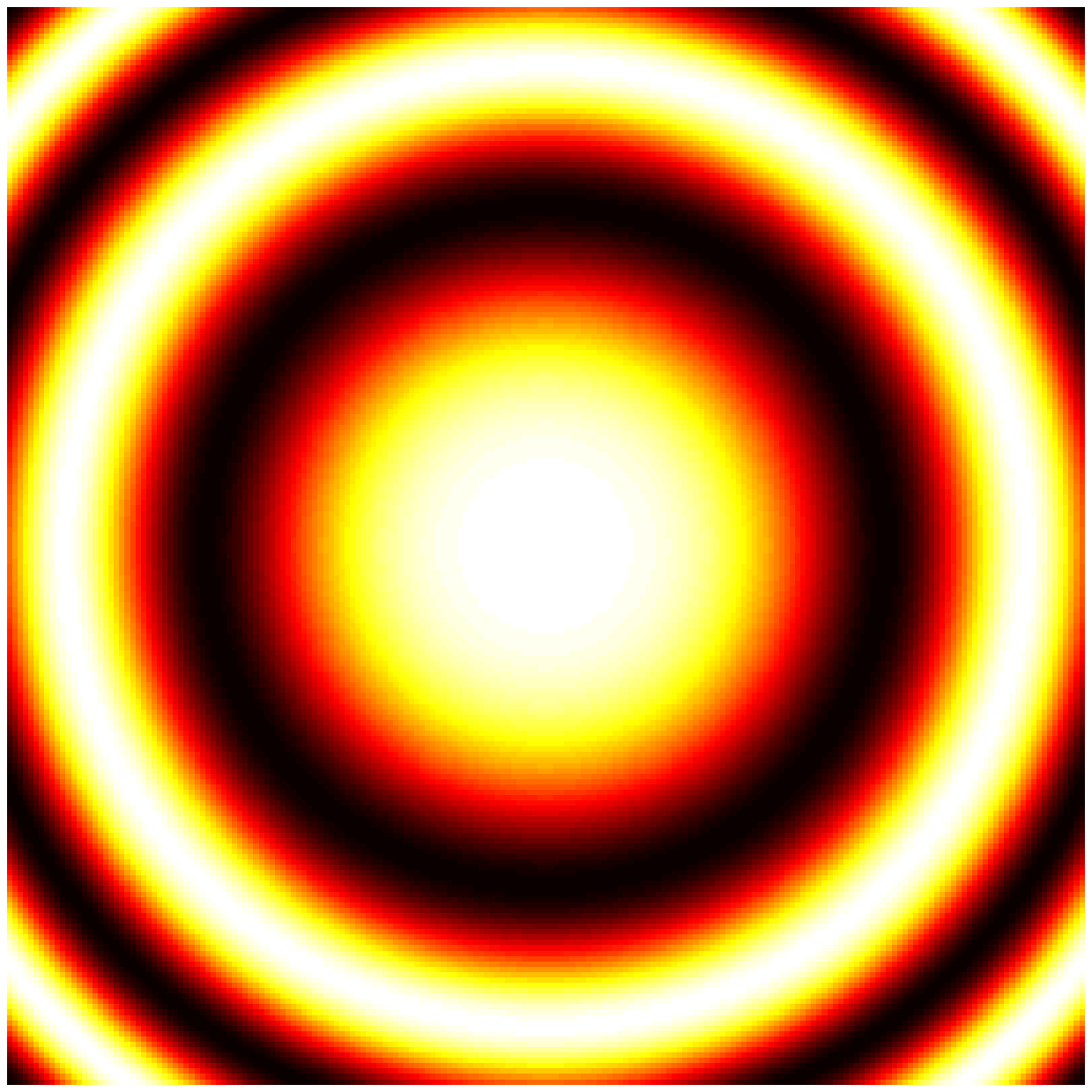} \\

\includegraphics[width=.22\textwidth]{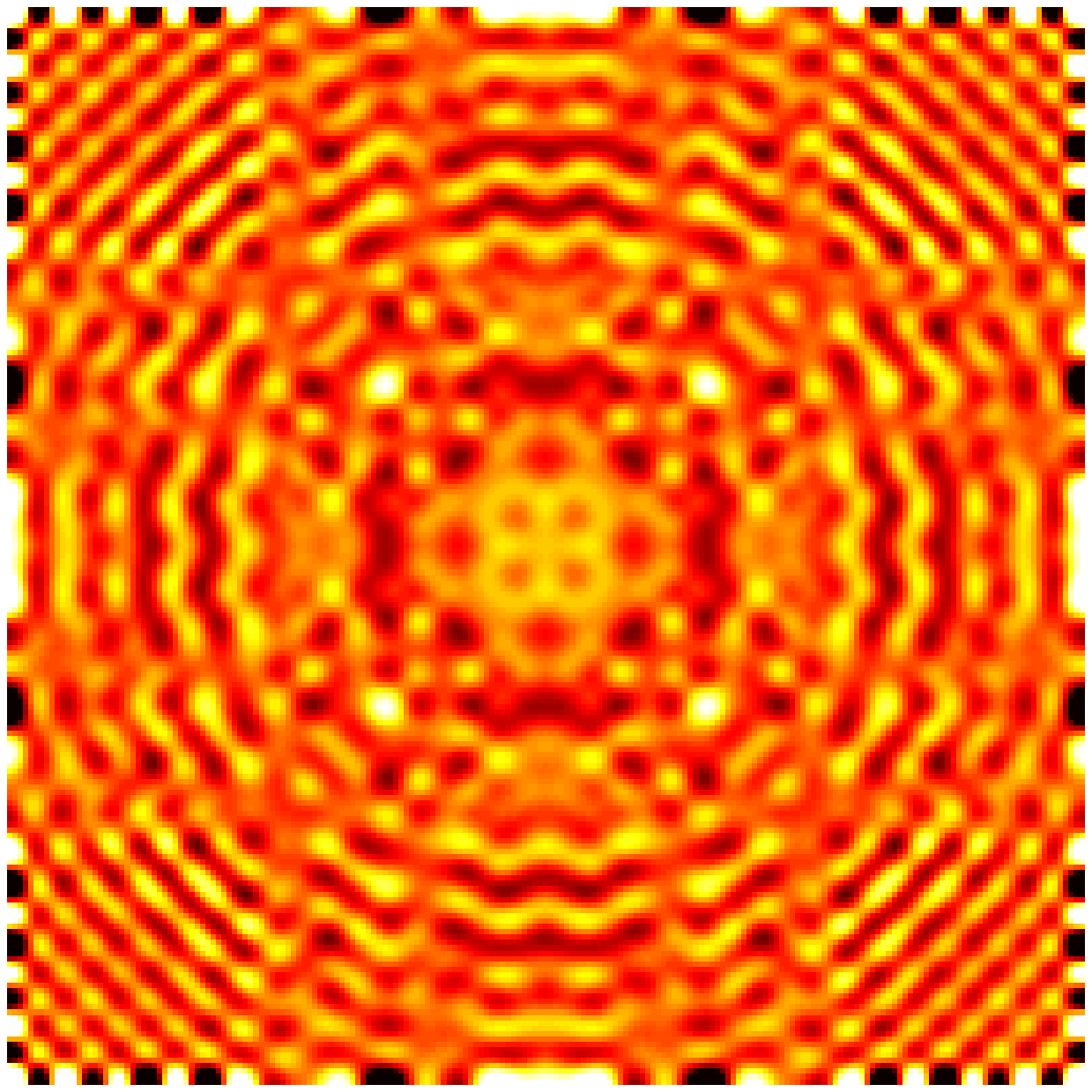} &
\includegraphics[width=.22\textwidth]{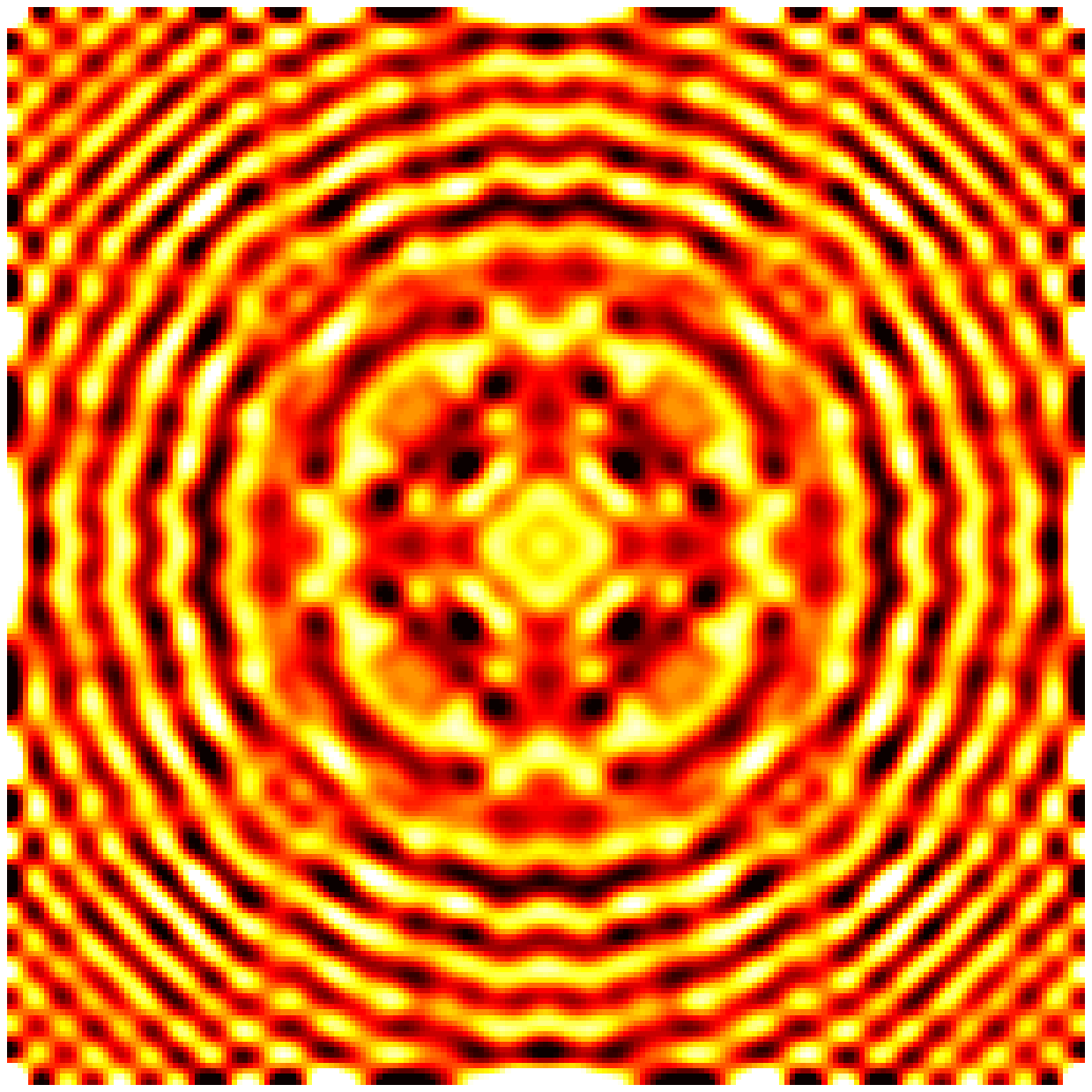} &
\includegraphics[width=.22\textwidth]{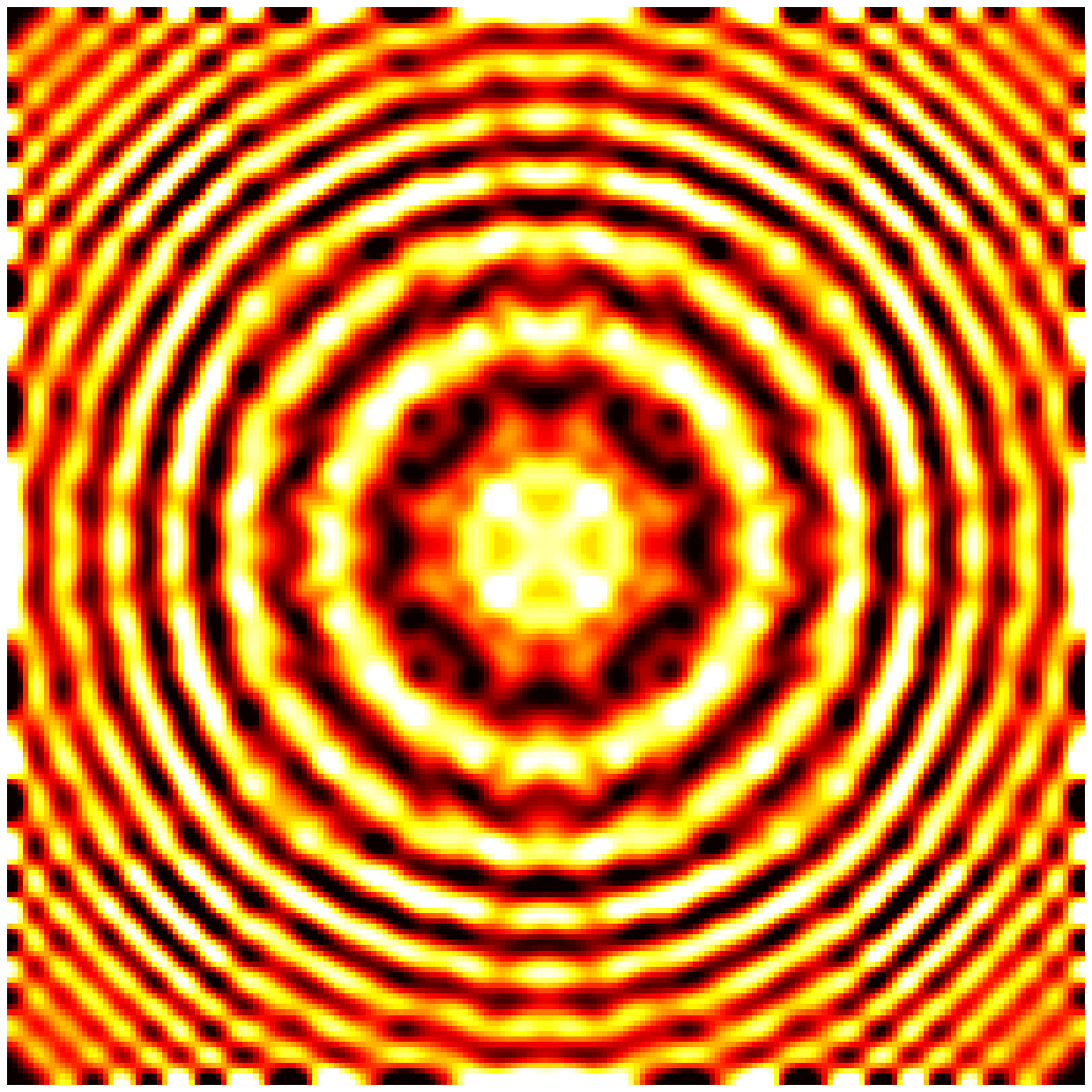} &
\includegraphics[width=.22\textwidth]{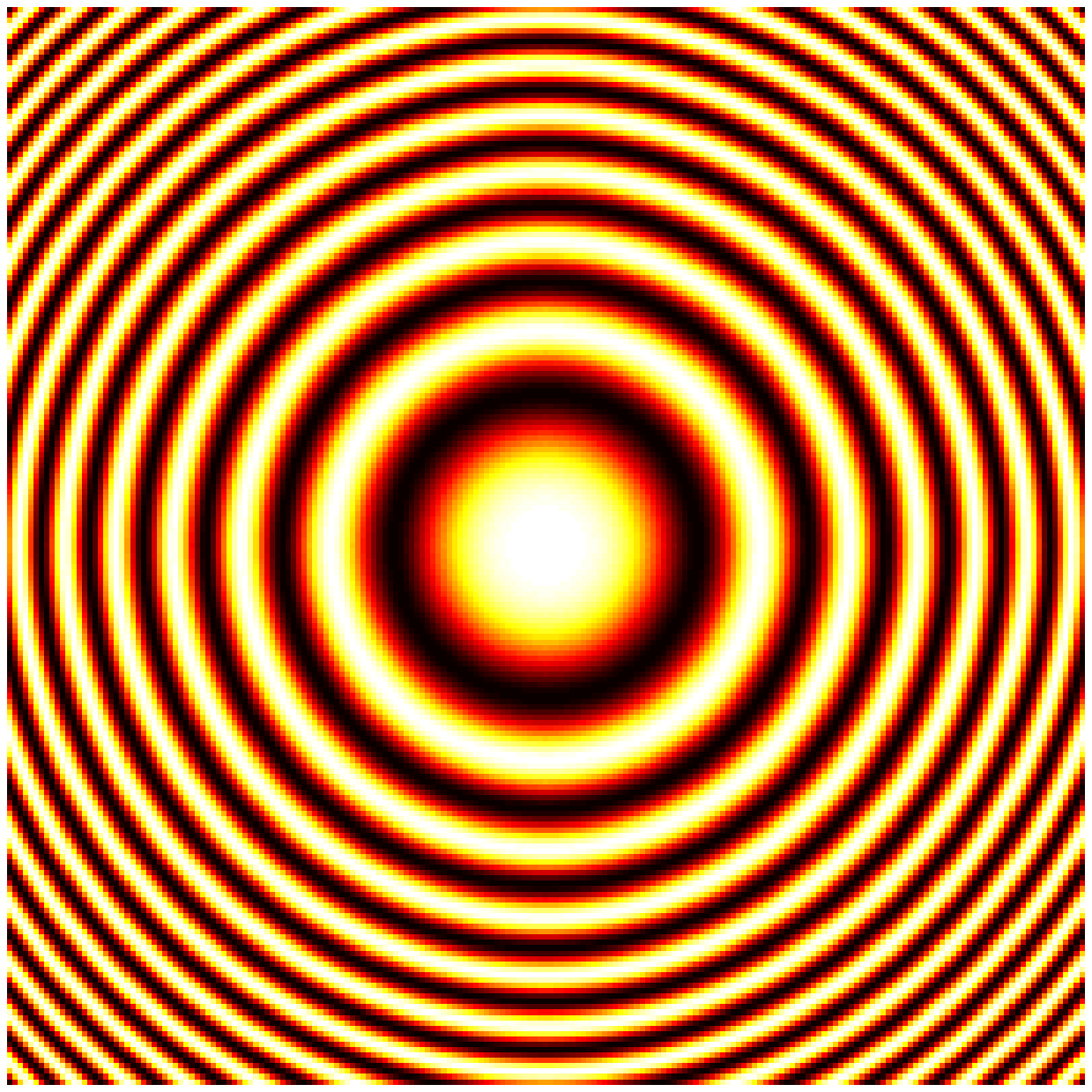} \\

\includegraphics[width=.22\textwidth]{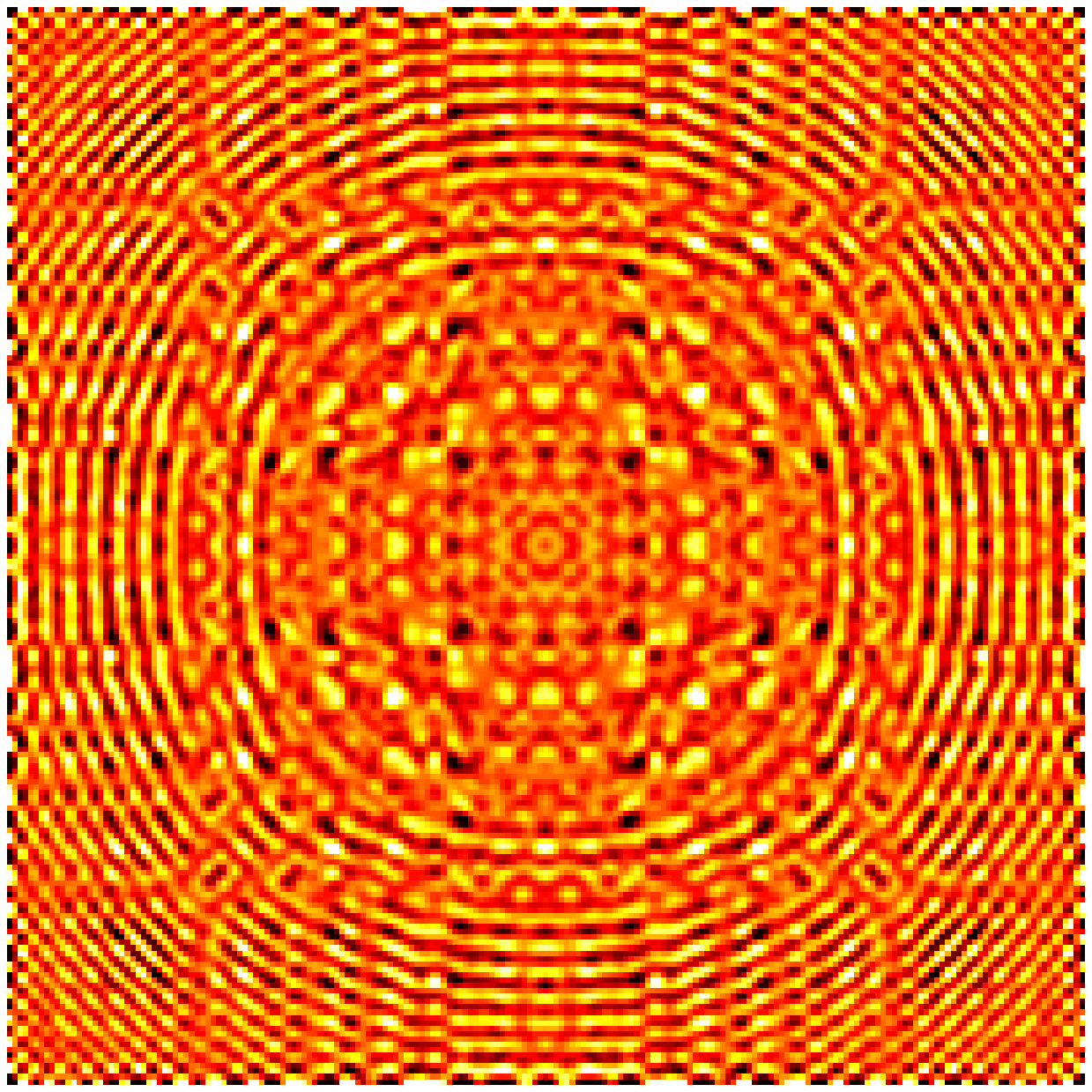} &
\includegraphics[width=.22\textwidth]{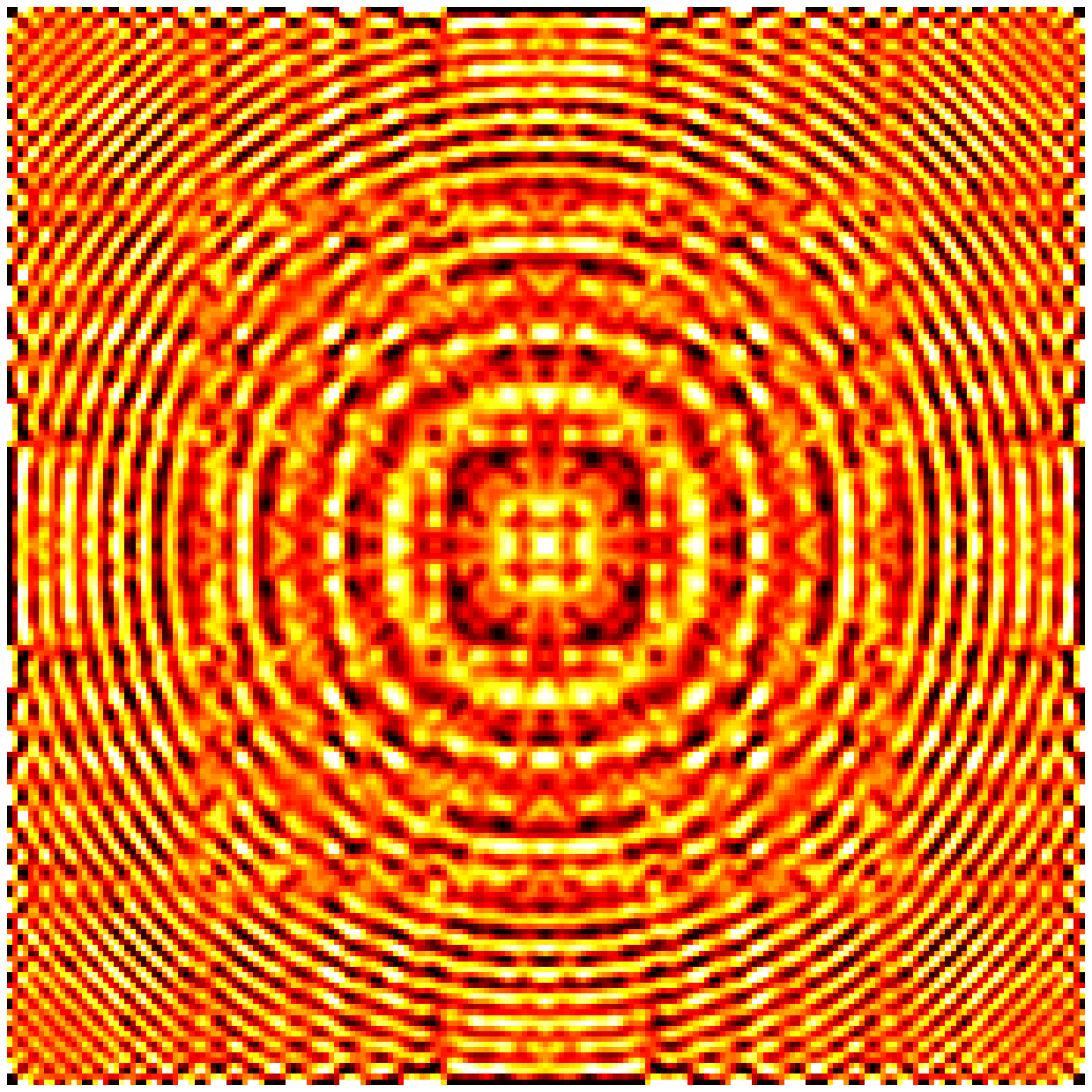} &
\includegraphics[width=.22\textwidth]{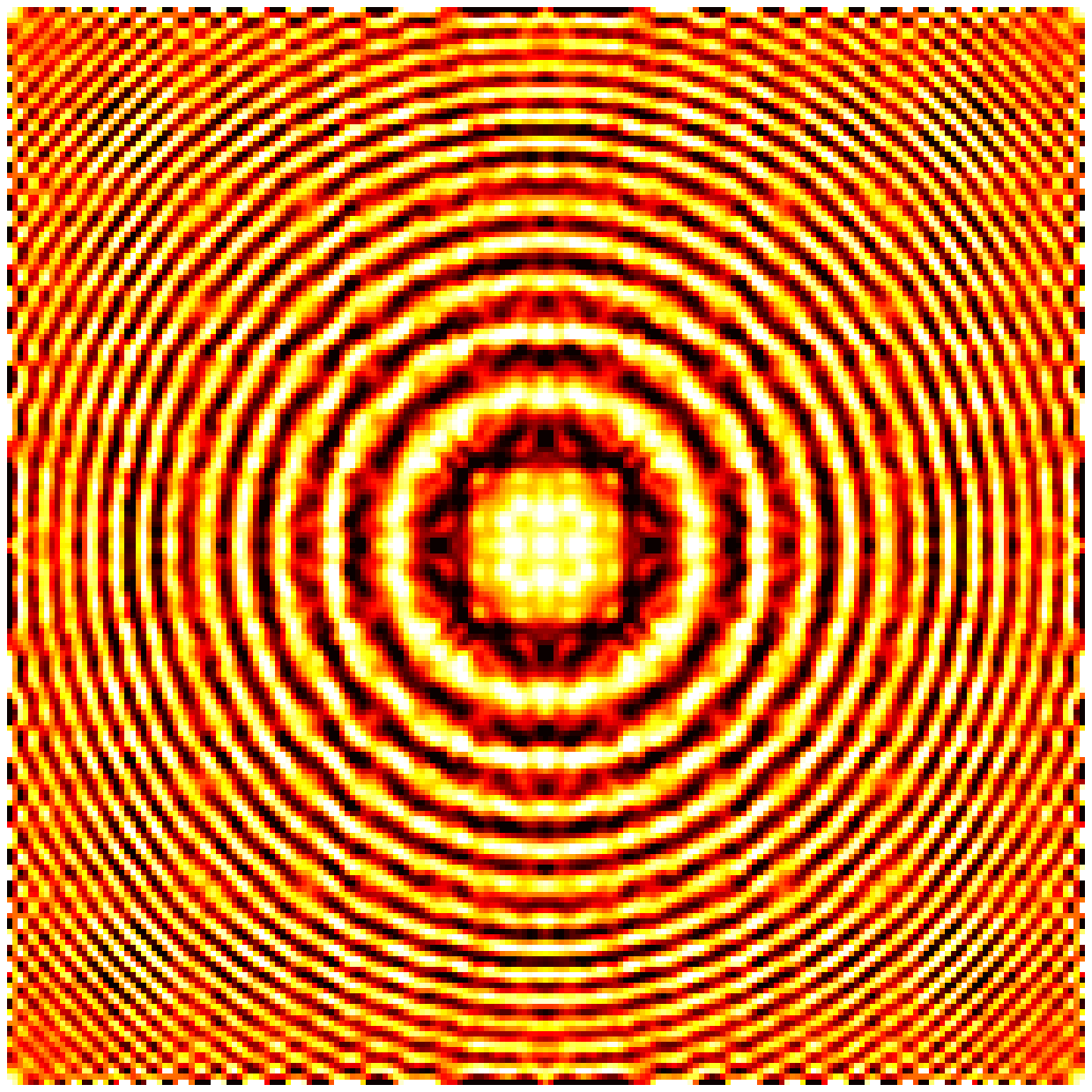} &
\includegraphics[width=.22\textwidth]{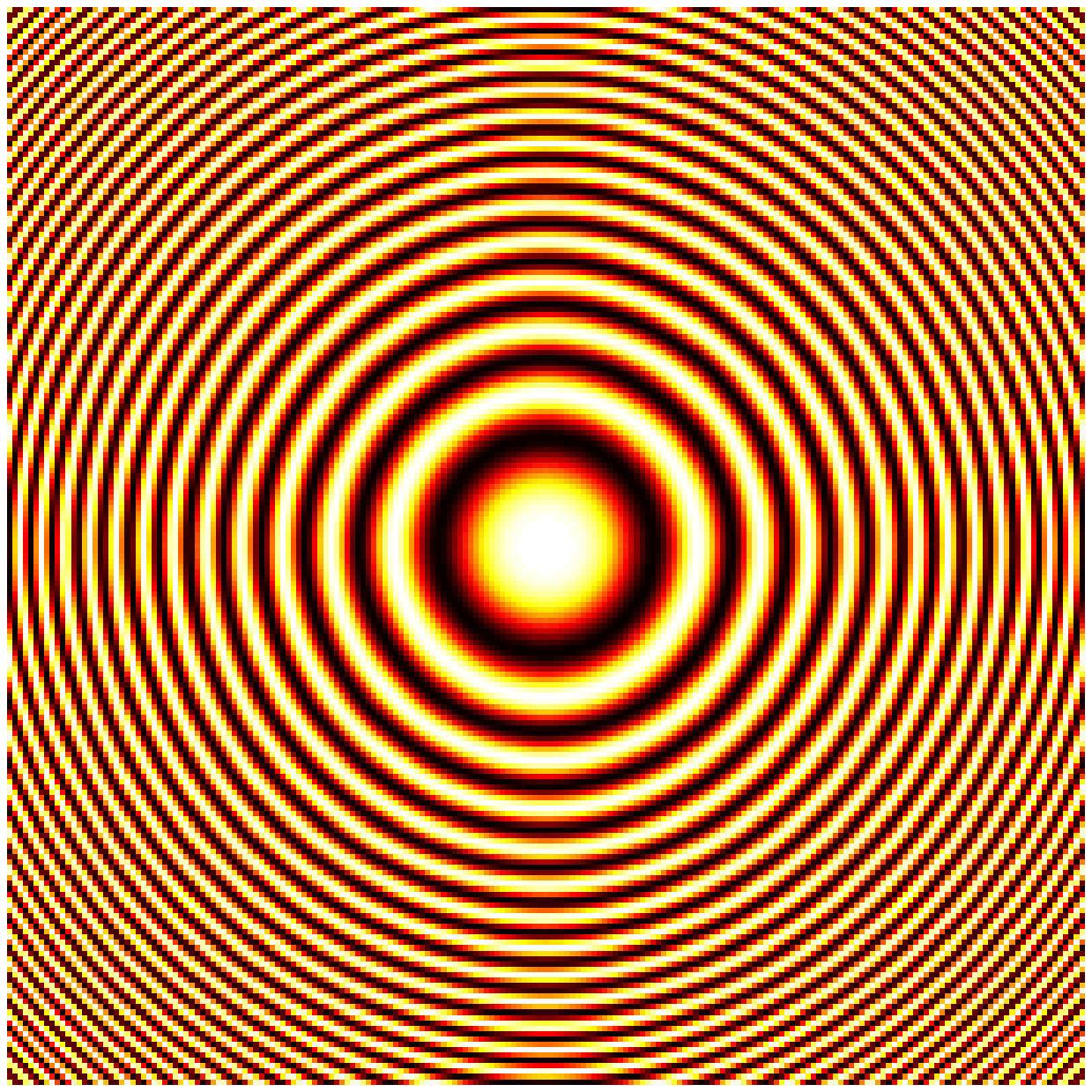} \\

\end{tabular}
\caption{The (real part of the) $w$-modulation kernel plotted in real space. For each row, a different $w_{\rm d}$ is considered: $w_{\rm d}=0.1$ for the upper row; $w_{\rm d}=0.5$ for the second row; and $w_{\rm d}=1.0$ for the lower row.  Notice the higher frequency content of the $w$-modulation kernel with increasing $w_{\rm d}$, \ie\ from highest to lowest row. For each column, a different approximation of the $w$-modulation kernel is plotted, for preserved energy levels $E=\{0.25, 0.50, 0.75, 1.00\}$, respectively, from the left to the right.  Notice that the kernels for $E<1$ more accurately approximate the exact kernel ($E=1$) as the energy preserved increases, \ie\ from left-to-right columns.}
\label{fig:chirp}
\end{figure*}

Following this strategy, we plot in Fig.~\ref{fig:chirp} examples of the \mbox{two-dimensional} $w$-modulation kernel in real space for different values of $w_{\rm d}$ and different preserved energy levels $E$. It can be clearly seen that a lower preserved energy proportion results in a greater distortion to the $w$-modulation kernel in real space. However, for a preserved  energy level of $E=0.75$ (shown in the third column of Fig.~\ref{fig:chirp}), the distortion is moderate. 

\begin{figure}
\includegraphics[width=0.47\textwidth, clip=true, trim= 0 0 0 0]{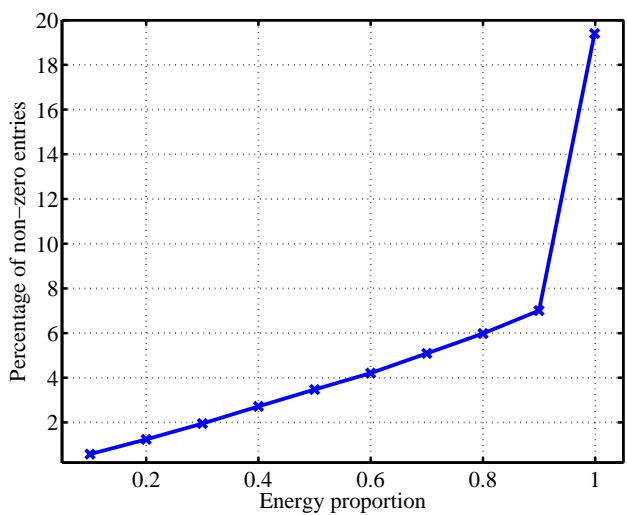}
\caption{Percentage of non-zero entries of the approximate $w$-modulation kernel $\hat{\mathbfss{C}}$ as a function of preserved energy proportion. Note that the point near $E=1$ is in fact $E=0.999$, highlighting the fact that $\hat{\mathbfss{C}}$ is sparse even when effectively no approximation is made.  Applying lower energy levels reduces the number of non-zero entries of $\hat{\mathbfss{C}}$ further.}
\label{fig:nnz}
\end{figure}

We study the relative sparsity of the sparse approximation of $\hat{\mathbfss{C}}$ by plotting in Fig.~\ref{fig:nnz} its proportion of non-zero entries. The size of the initial matrix operator $\hat{\mathbfss{C}}$ with dimensions $M \times N^2$ corresponds to 100\pc\ of non-zero entries.  Sparsifying the matrix with a preserved energy of $E=0.999$ already decreases the proportion of non-zero entries to $\sim$20\pc. This case effectively corresponds to no approximation and shows the sparse nature of the exact $w$-modulation kernel in Fourier space. Applying lower energy levels reduces the number of non-zero entries of $\hat{\mathbfss{C}}$ further, to 7\pc\ for $E=0.9$ and to 0.6\pc\ for $E=0.1$.

In the following, we address the question to what degree the approximation degrades reconstruction quality. In order to do so, we simulate full image reconstructions. The experimental set-up of these simulations is described in detail in Sec.~\ref{sec-sim}, where a thorough study of the effect of $w$-modulation is performed.  Here we study the impact of the sparse approximation of the matrix operator $\hat{\mathbfss{C}}$ only, in order to select a suitable preserved energy proportion $E$ for the subsequent analysis performed in Sec.~\ref{sec-sim}.

\begin{figure}

\includegraphics[width=0.47\textwidth, clip=true, trim= 0 0 0 0]{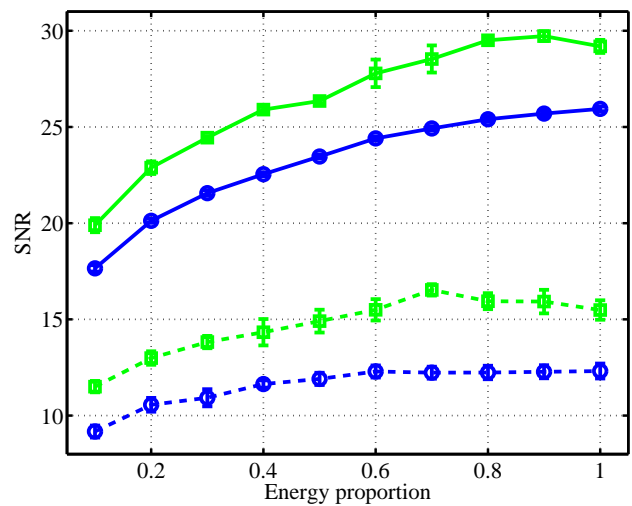}
\caption{Reconstruction quality of M31 (green lines marked with squares) and 30Dor (blue lines marked with circles) as a function of preserved energy proportion of the approximate $w$-modulation operator $\hat{\mathbfss{C}}$.  Daubechies 8 wavelets are used as the sparsity basis, while a visibility coverage of 10\pc\ is considered in the dashed lines and a coverage of 50\pc\ is considered in the solid lines (see text of Sec.~\ref{sec-sim} for further details).  The data points are the mean of 10 simulations and the error bars show one-standard-deviation.  Notice that the level of preserved energy can be reduced considerably without markedly reducing reconstruction quality.}
\label{fig:energySNR}
\end{figure}

We vary the preserved energy level $E$ used to construct the sparse approximation of $\hat{\mathbfss{C}}$ and examine the image reconstruction quality of the two test images shown in Fig.~\ref{fig:original} (these images and the experiment set-up will be introduced in detail in Sec.~\ref{subsec-setup}). We measure the effectiveness of the sparse matrix approximation of $\hat{\mathbfss{C}}$ through the signal-to-noise ratio (SNR) of the reconstructed image (defined explicitly in Sec.~\ref{subsec-setup}) and the runtime of image reconstruction relative to the runtime of a full energy reconstruction, both as a function of the preserved energy level $E$.
The SNR of the reconstructed images as a function of preserved energy proportion is shown in Fig.~\ref{fig:energySNR} for visibility coverages of 10\pc\ and 50\pc, where $w$-components vary. It can be seen clearly that the quality of reconstruction remains close to the case where no sparsification is made ($E=1$) for relatively low energy proportions.  Below $E\sim0.5$ for 10\pc\ coverage and below $E\sim0.7$ for 50\pc\ coverage, reconstruction quality slowly starts to decrease.

\begin{figure}
\centering
\includegraphics[width=0.47\textwidth, clip=true, trim= 0 0 0 0]{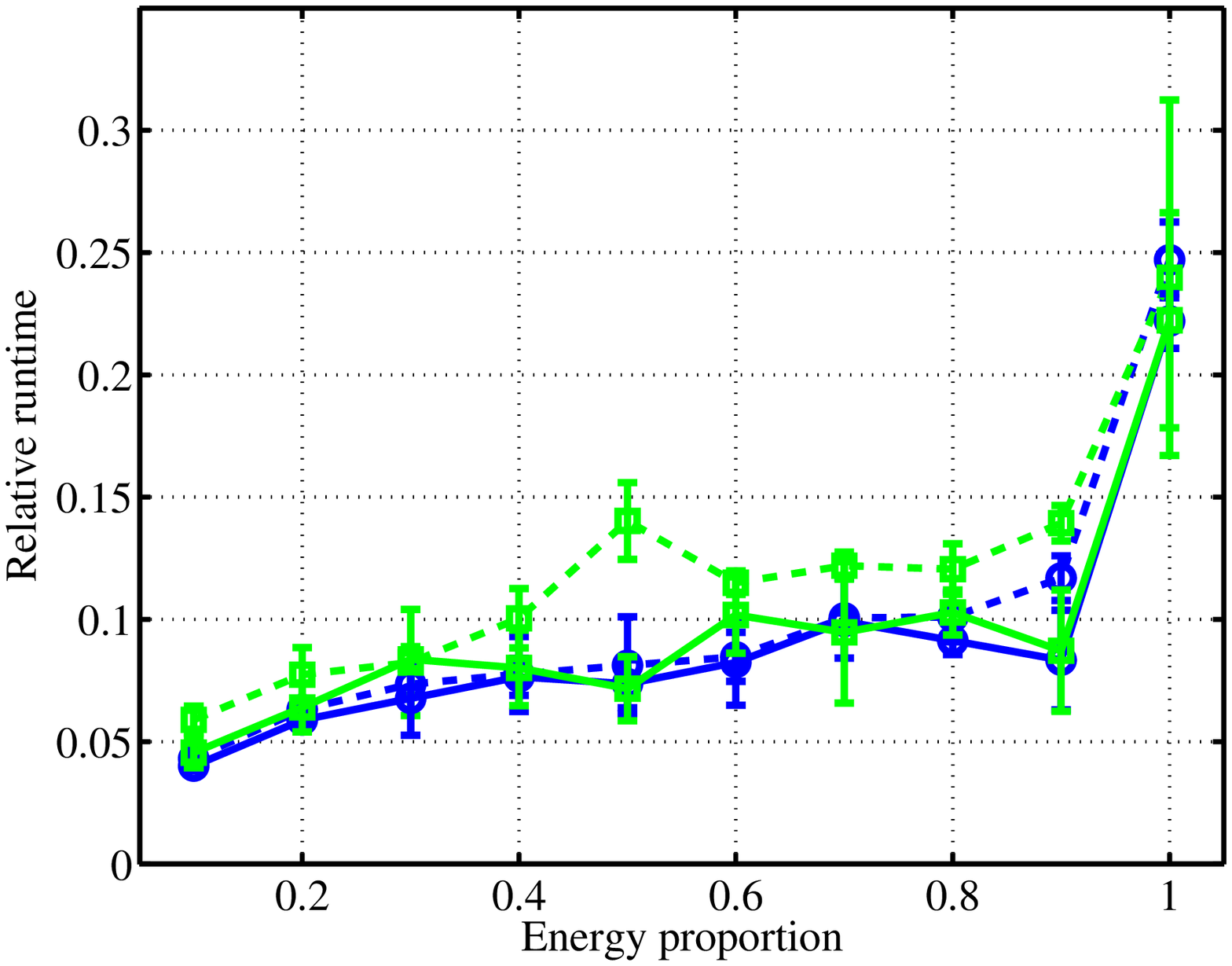}

\caption{Relative computation-runtime as a function of the preserved energy with visibility coverage of 10\pc\ in the dashed lines and a coverage of 50\pc\ in the solid lines. The green curves refer to the simulations using M31 (marked with squares) and the blue curves to 30Dor (marked with circles). The data points are the mean of 10 simulations and the error bars show one-standard-deviation.  The point close to $E=1$ is in fact $E=0.999$, highlighting the fact that the natural sparsity of $\hat{\mathbfss{C}}$ can be exploited even when effectively no approximation is made. Notice that by reducing the level of preserved energy, runtime can be reduced considerably.}
\label{fig:energytime}
\end{figure}

The relative runtime of image reconstruction implemented in MATLAB and run on a state-of-the-art 6 core machine is plotted in Fig.~\ref{fig:energytime}.  It can be seen clearly that the runtime is significantly decreased by reducing $E$. For example, for M31 with 10\pc\ visibility coverage, the computation-time is already reduced from 4h12min for $E=1$ to 60min for $E=0.999$ (where effectively no approximation of $\hat{\mathbfss{C}}$ is made).  Reducing the preserved energy proportion further, computation-time is reduced to 29min for \mbox{$E=0.9$}, and to 21min for $E=0.6$ where reconstruction quality is still not markedly degraded. 

The computational cost of image reconstruction is related closely to the asymptotic complexity of application of the measurement operator, which we consider next for various algorithms.  In the case each visibility has a different $w$-component, if the $w$-projection algorithm is not applied, then there is no advantage in applying an FFT.  In this setting one would apply a discrete Fourier transform; thus, application of the measurement operator would be of order $\mathcal{O}(M N^2)$.  If $w$-projection is applied, using an FFT but without any sparsification of $\hat{\mathbfss{C}}$, then the complexity becomes $ \mathcal{O}(M N^2 + N^2 \log N)$, where the first term corresponds to multiplication by a dense matrix $\hat{\mathbfss{C}}$ of size $M \times N^2$ and the second to the FFT.  By sparsifying $\hat{\mathbfss{C}}$, as described above, then the complexity is reduced to $\mathcal{O}(S + N^2 \log N)$, where $S$ specifies the number of non-zero entries of $\hat{\mathbfss{C}}$.  Sparsifying $\hat{\mathbfss{C}}$ can thus reduce the computational cost of imaging considerably, as also demonstrated by the numerical experiments performed above.

In our subsequent analysis, we take an extremely conservative approach and select a preserved energy level of $E=0.75$.  This is sufficient to render a study of the spread spectrum effect for varying $w$-component computationally tractable in the experimental setting described in Sec.~\ref{subsec-setup}, while we can also be confident that reconstruction quality is not degraded.  For future applications, the preserved energy proportion could be further reduced, provided that the implications for image reconstruction are considered carefully.

 \section{Advantage of $w$-components: Simulations and results}
\label{sec-sim}

In this section we study the impact of the spread spectrum effect for various $w$-modulation scenarios, including constant and varying $w$-components, using extensive numerical simulations.  Firstly, the experimental parameters of our simulations are outlined.  Secondly, image reconstruction results are presented and discussed for the different $w$-modulation scenarios, where we also consider different images and choices of sparsifying dictionaries.

\subsection{Experimental set-up}
 \label{subsec-setup}
\begin{figure}
\centering
\subfigure[M31]{         
                  \centering  
                \includegraphics[width=.38\textwidth, clip=true, trim= 0 30 0 30]{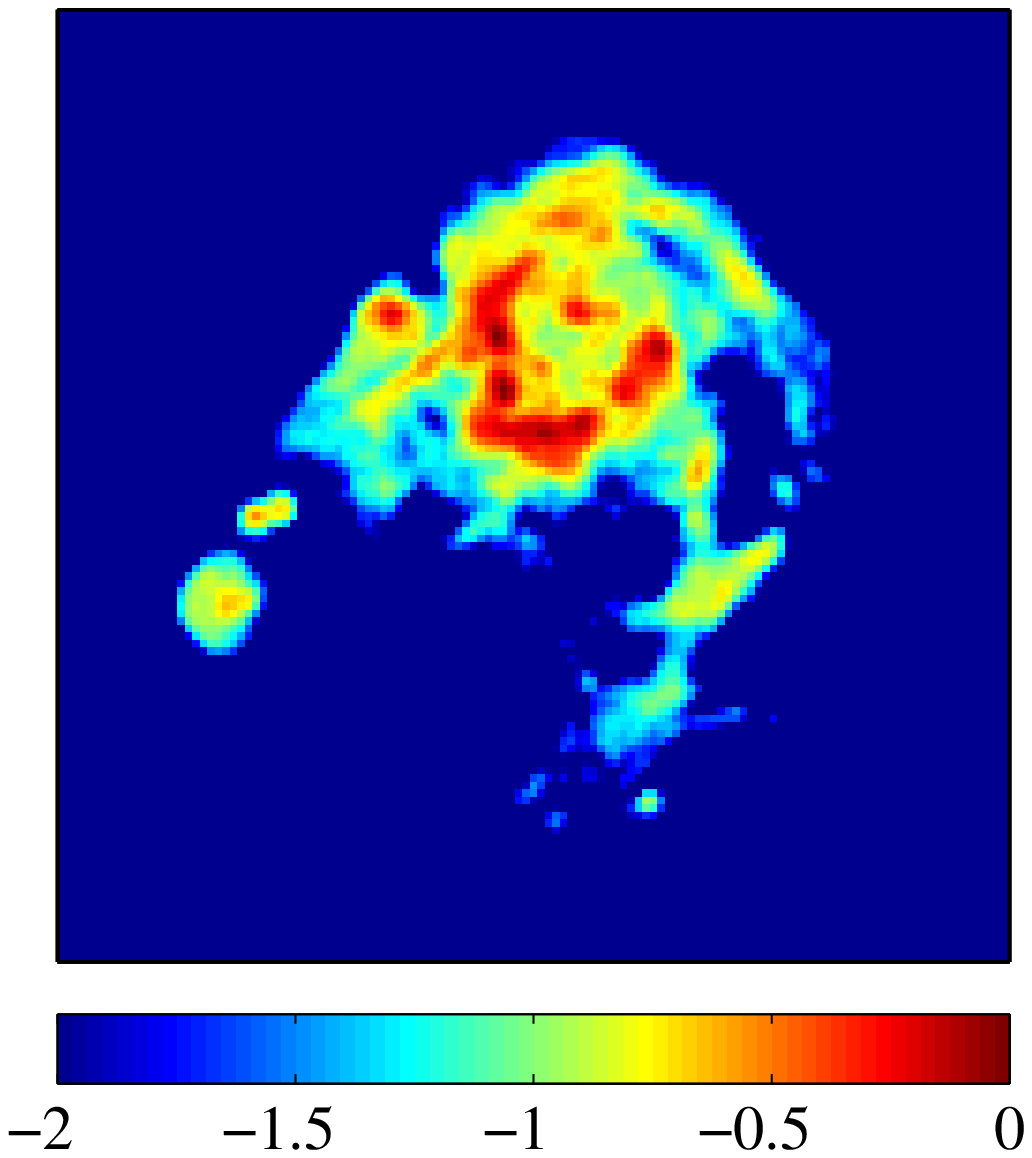}
                \label{fig:M31}
        }
        
        \subfigure[30Dor]{         
                  \centering  
                \includegraphics[width=.38\textwidth, clip=true, trim= 0 30 0 30]{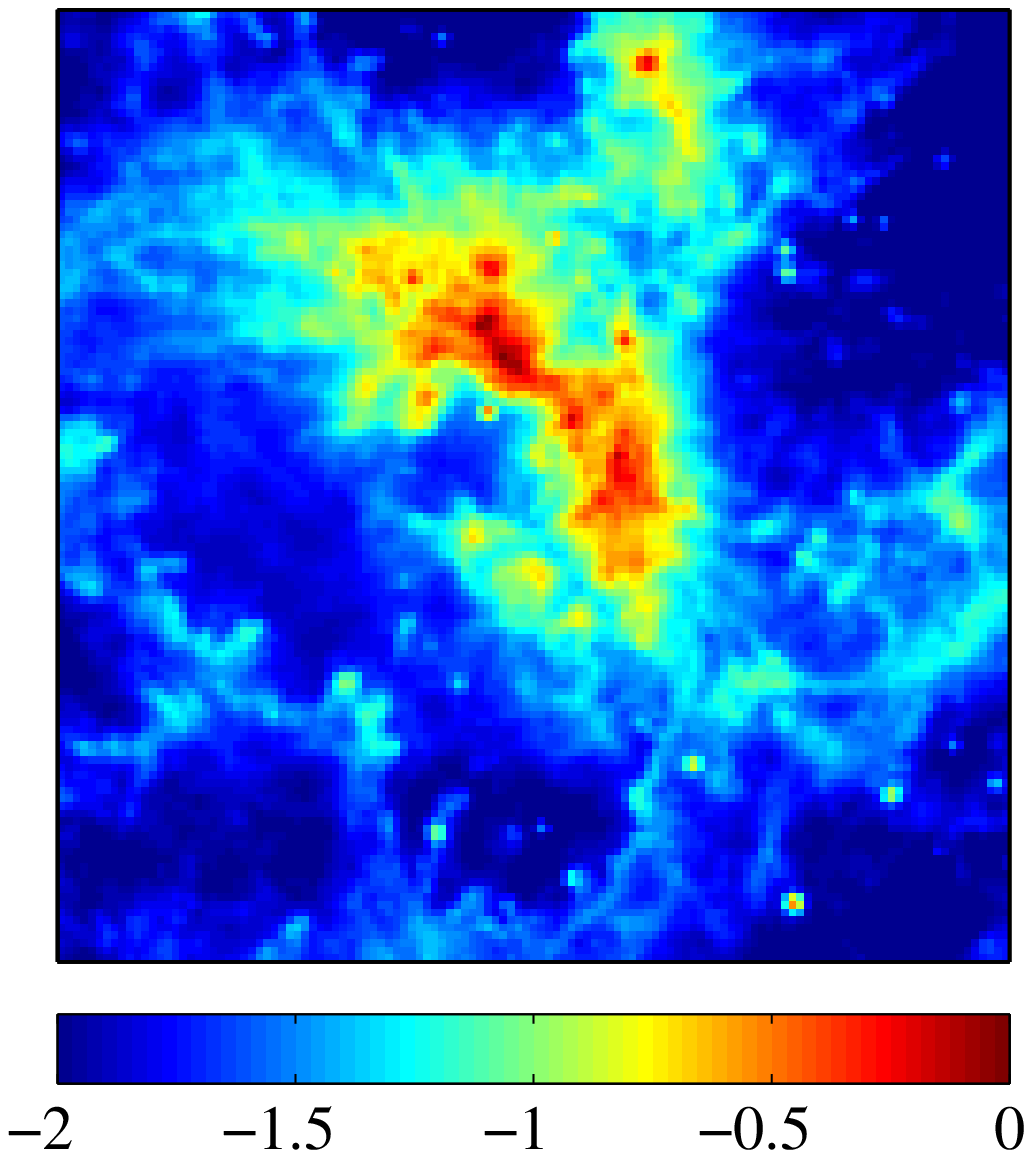}
                \label{fig:30dor}
        }

\caption{Original images in logarithmic scale.}
\label{fig:original}
\end{figure}

We investigate the effect of $w$-modulation on reconstruction quality using two test images: an HII region in M31 and 30 Doradus (30Dor) in the Large Magellanic Cloud, as shown in Fig.~\ref{fig:original}. We choose two images to demonstrate reconstruction quality on images with different spatial distribution characteristics. For computational reasons, we downsample the original test images\footnote{Available at \url{http://casaguides.nrao.edu/index.php}.} (yielding an $N \times N = 125\times 125$ pixel image for M31 and an  $N \times N = 127\times 127$ pixel image for 30Dor).

The opening angle  of the FoV is chosen constant throughout our analysis  at $\theta=2^\circ$, with corresponding planar FoV \mbox{$L=2\sin(\frac{\theta}2)$}.  Note, however, that our analysis is insensitive to the choice of $\theta$ since different choices just rescale the visibility coordinates $(u,v,w)$.

The band-limit $B$ of a pixelated image is related to its FoV $L$ and the number of pixels along each dimension $N$ via $B=\frac N{2L}$. The band-limit of the linear chirp approximation of the $w$-modulation, as specified by Eq.~(\ref{equ-chirp1}), is given approximately by its instantaneous frequency: $B^{\rm{Chirp}}\simeq w_{\rm d} \frac{N}{2L} =w_{\rm d} B$, where we introduce the variable $w_{\rm d}=w\frac{L^2}{N}$ \citep{Wiaux:2009eh,McEwen:2010cq}.  As commented already, the $w$-modulation can be seen as a convolution in Fourier space; thus, the band-limit of the convolved image is given by $B+B^{\rm Chirp} \simeq (1+w_{\rm d})B$.  We consider the maximal case of $w_{\rm d}=1$, where the band-limit is doubled.  In order to support the spreading due to the maximal $w$-component, it is necessary to double the band-limit, which requires doubling the pixel support of the image.  Upsampling of the image by a factor of two is applied, in order to support this increased band-limit, by the application of the operator $\mathbfss{U}$, which implements zero-padding in Fourier space (as described in Sec.~\ref{sec-ri-op}). In this scenario, the maximal $w$-component considered is $\frac{2}{L}$ larger than the maximal $u$ or $v$ component considered. 

We sample the visibilities on a $2N \times 2N$ grid and reconstruct images from visibility coverages ranging from 10\pc\ to 90\pc\ of the full-set of discrete visibilities. We assume uniformly random sampling distributions in the visibility coordinates $(u,v,w)$, in order to disregard considerations due to particular telescope configurations and to perform large numbers of simulations.

We simulate instrumental noise $ \bmath n$ under the assumption of independent identically distributed Gaussian noise. The width of the Gaussian distribution is calculated via $\sigma(\bmath n)=10^{-\rm{SNR_{in}}/20}\sigma(\bmath x)$, where $\sigma(\bmath x)$ is the standard deviation of the input image. The input signal-to-noise ratio $\rm{SNR_{in}}$ is chosen as $30$ dB throughout the article. Reconstruction quality is measured by the $\rm{SNR_{out}}$ of the recovered image, calculated via \mbox{$\rm{SNR_{out}}=20\log_{10}(\sigma(\bmath x)/\sigma(\bmath x- \bmath x_{\rm{rec}}) )$}, where $\sigma(\bmath x-\bmath x_{\rm{rec}})$ is the standard deviation of the difference of original and recovered image. Our results are shown in units of dB and can be easily compared to the input noise level of $30$ dB.

\subsection{Results}

We reconstruct images from simulated visibility measurements for various $w$-modulation scenarios: no $w$-component (\ie\ constant $w_{\rm d}=0$); a uniformly random varying $w$-component (with $w_{\rm d} \sim \mathcal{U} (0, 1)$); and a constant, maximal $w$-component (\ie\ constant $w_{\rm d}=1$). Our main focus is to study the more realistic varying $w$-component scenario, since the impact of the spread spectrum effect for constant $w$-component has been studied by \cite{Wiaux:2009eh}, where the improvement in reconstruction quality due to a non-zero $w$-modulation was demonstrated already.  Furthermore, we also vary the visibility coverage, image and sparsity dictionary, in order to demonstrate the robustness of our results. The presented results are simulated using an energy threshold of $E=0.75$ for the sparse approximation of the $w$-modulation operator.

In our analysis we consider the Dirac and Daubechies 8 wavelet sparsity bases, and also the SARA algorithm \citep{Carrillo:2012uv}. Reconstruction with the Dirac basis can be considered as a proxy for the CLEAN algorithm \citep{hogbom74}; the close correspondence was shown previously by \cite{Wiaux:2008ed}.  Daubechies 8 wavelets can be considered as a proxy for multi-scale CLEAN \citep{Cornwell:2008zn}; the Daubechies 8 wavelet reconstruction here can in fact be considered as an upper bound on the reconstruction quality of multi-scale CLEAN (since \cite{Li:2011wc} showed $\ell_1$ reconstruction using the isotropic undecimated wavelet transform to be superior to multi-scale CLEAN using M31 as a test image, and \cite{Carrillo:2012uv} in turn showed Daubechies 8 wavelets to be superior to the isotropic undecimated wavelet transform for both the test images used in this work).  We also consider the SARA algorithm developed recently by \cite{Carrillo:2012uv}, which exhibits reconstruction performance substantially superior to both CLEAN- and multi-scale CLEAN-like algorithms. 

\begin{figure*}       
          \includegraphics[width=1.\textwidth, clip=true, trim= 50 55 0 100]{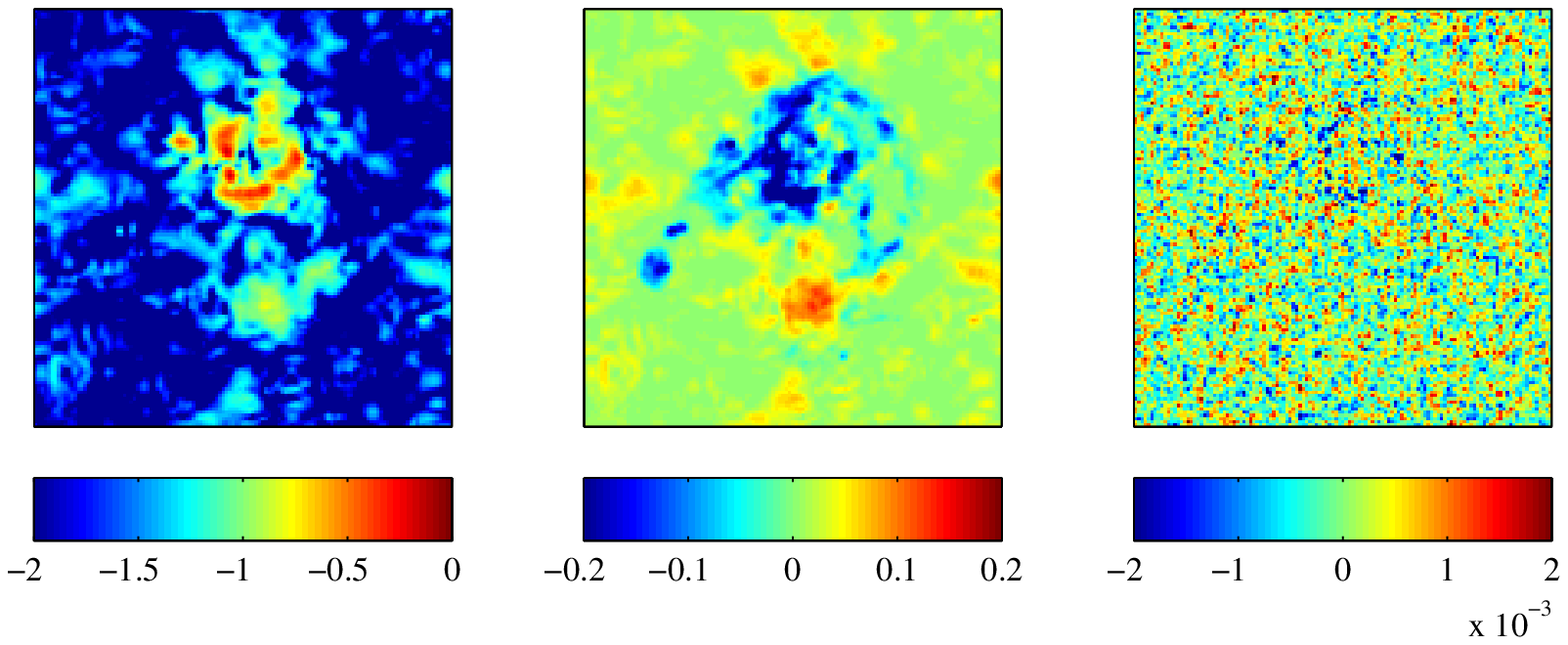}\\
          \includegraphics[width=1.\textwidth, clip=true, trim= 50 55 0 100]{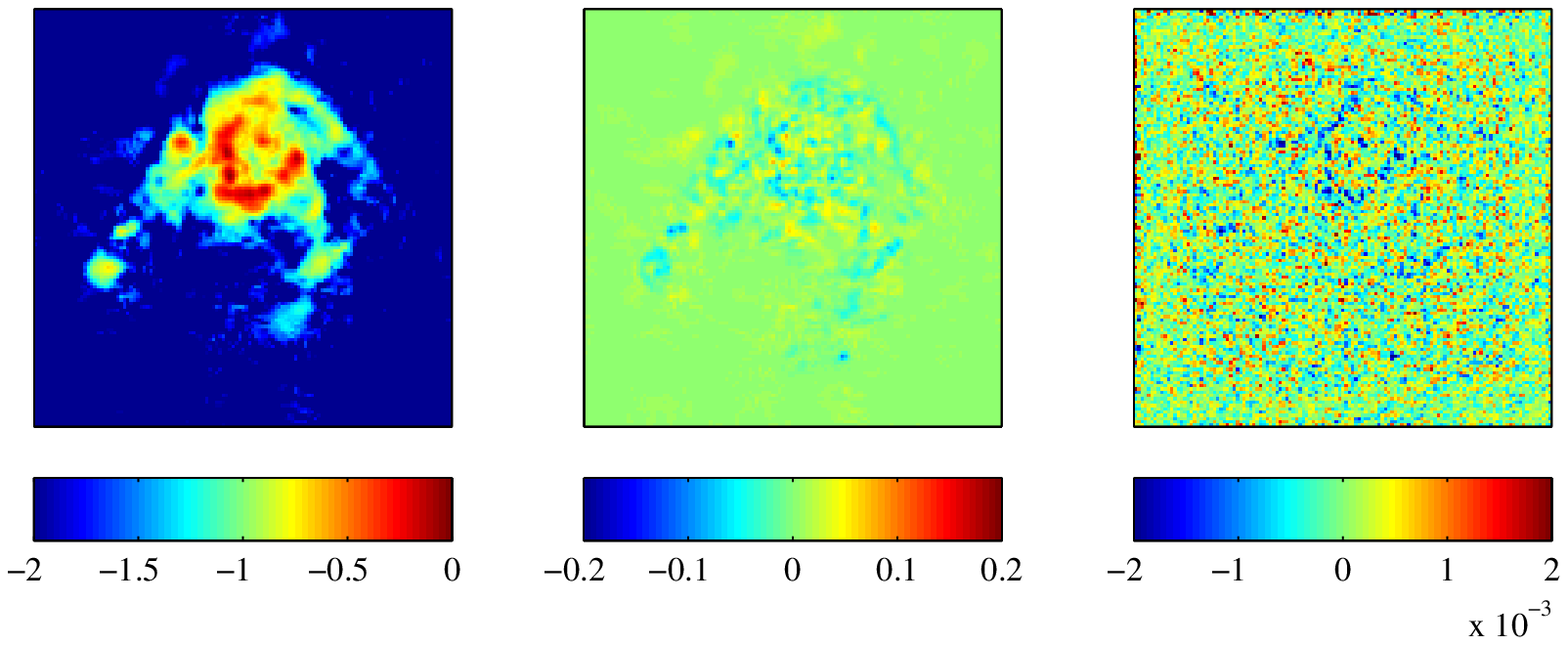}\\
          \includegraphics[width=1.\textwidth, clip=true, trim= 50 0 0 100]{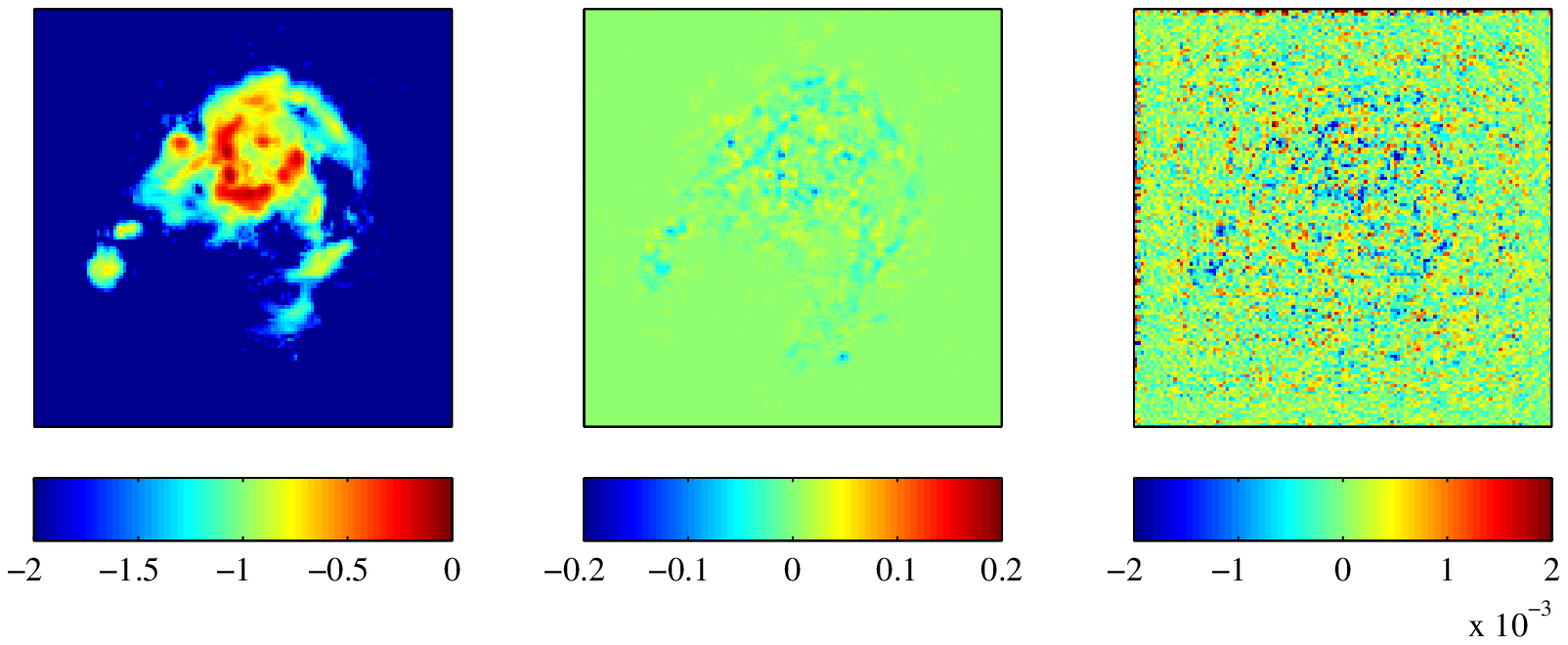}
          \caption{Image reconstruction of M31 using Daubechies 8 wavelets with visibility coverage of 10\pc. The input noise level is 30 dB and the energy threshold is chosen as $E=0.75$. Each row shows a reconstructed image in $\log_{10}$ scaling (left panel), and in linear scaling the difference of the original and the reconstructed image (middle panel) and the residual image (right panel). The first row corresponds to no $w$-modulation (\ie\ $w_{\rm d}=0$), the second one to the uniformly random $w$-component case (\ie\ $w_{\rm d} \sim \mathcal{U}(0,1)$), and the third row to maximal, constant $w$-component (\ie\ $w_{\rm d}=1$). The $\rm{SNR_{out}}$ of the reconstructed images is, respectively, $\rm{SNR_{out}}=5$ dB for no $w$-modulation, $\rm{SNR_{out}}=16$ dB for varying $w$-component, and $\rm{SNR_{out}}=19$ dB for maximal, constant $w$-component.  Clearly, the quality of reconstruction for the varying $w$-component scenario (second row) is very close to the reconstruction quality for maximal, constant $w$-component (third row). }
          \label{fig:imM31}
        \end{figure*}

\begin{figure*}       
          \includegraphics[width=1.\textwidth, clip=true, trim= 50 55 0 100]{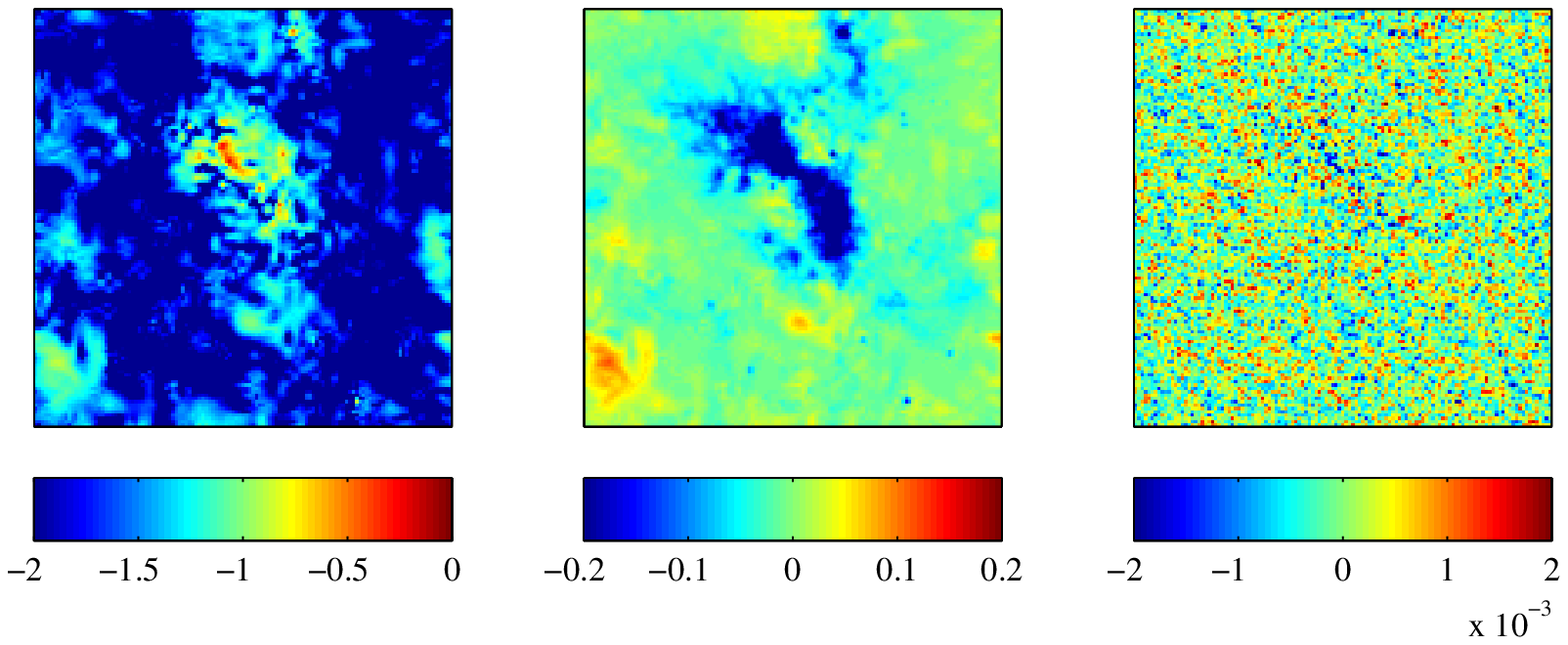}\\
          \includegraphics[width=1.\textwidth, clip=true, trim= 50 55 0 100]{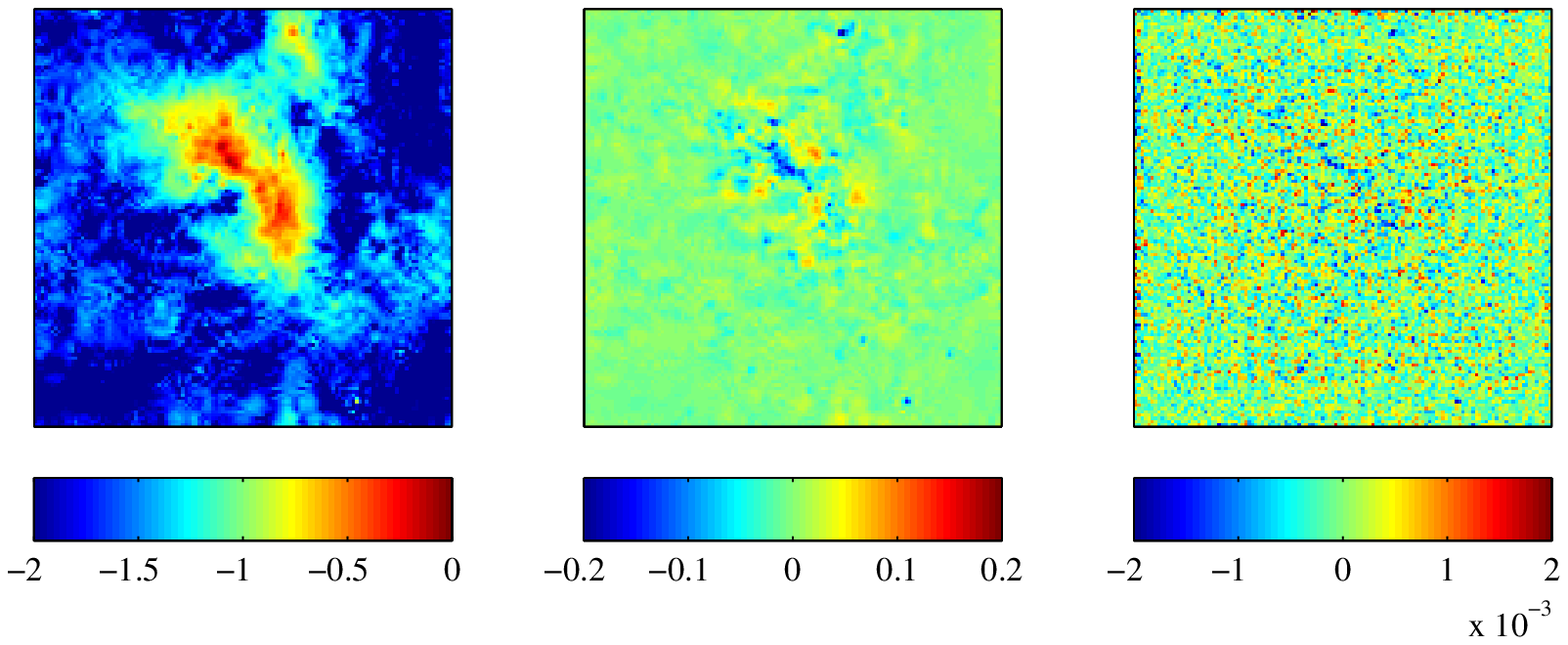}\\
          \includegraphics[width=1.\textwidth, clip=true, trim= 50 0 0 100]{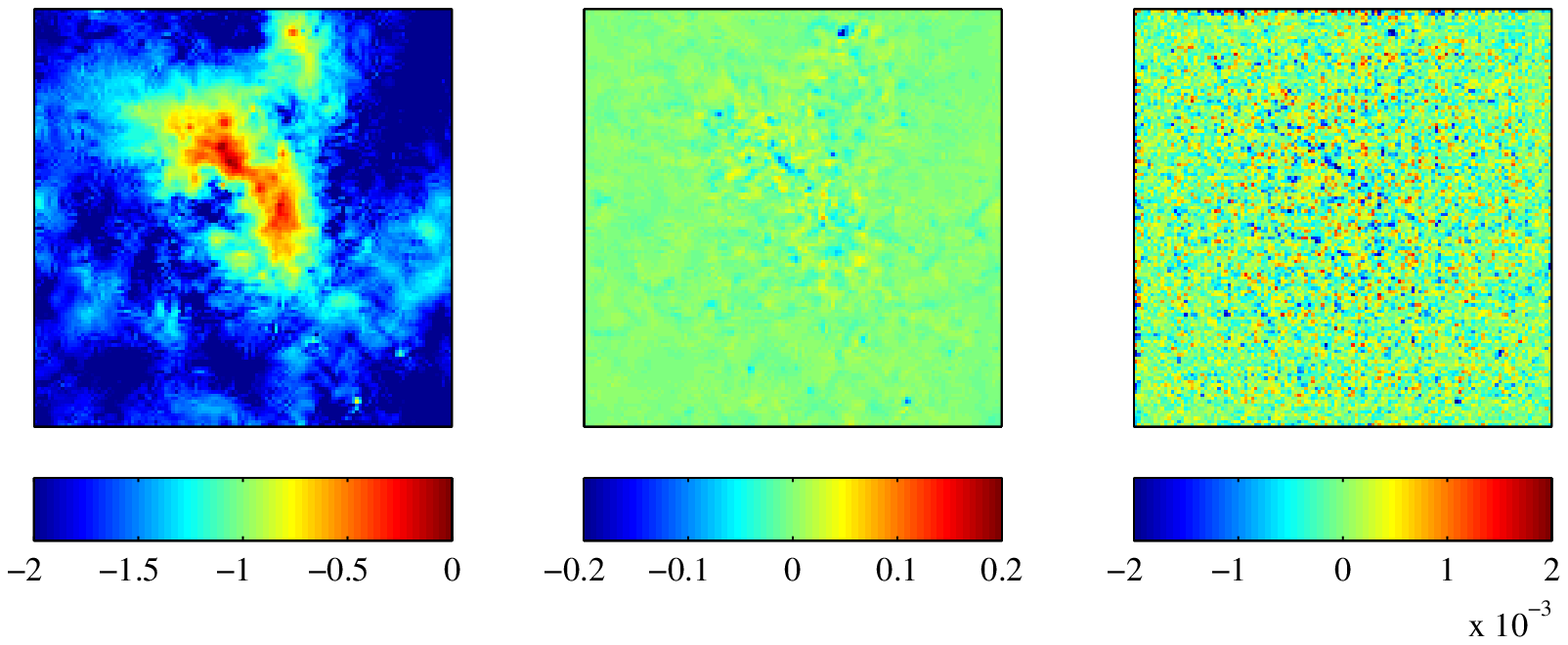}
          \caption{Image reconstruction of 30Dor using Daubechies 8 wavelets with visibility coverage of 10\pc. The input noise level is 30 dB and the energy threshold is chosen as $E=0.75$. Each row shows a reconstructed image in $\log_{10}$ scaling (left panel), and in linear scaling the difference of the original and the reconstructed image (middle panel) and the residual image (right panel). The first row corresponds to no $w$-modulation (\ie\ $w_{\rm d}=0$), the second one to the uniformly random $w$-component case (\ie\ $w_{\rm d} \sim \mathcal{U}(0,1)$), and the third row to maximal, constant $w$-component (\ie\ $w_{\rm d}=1$). The $\rm{SNR_{out}}$ of the reconstructed images is, respectively, $\rm{SNR_{out}}=2$ dB for no $w$-modulation, $\rm{SNR_{out}}=12$ dB for varying $w$-component, and $\rm{SNR_{out}}=15$ dB for maximal, constant $w$-component.  Clearly, the quality of reconstruction for the varying $w$-component scenario (second row) is very close to the reconstruction quality for maximal, constant $w$-component (third row). }
          \label{fig:im30dor}
        \end{figure*}

Firstly, we provide a qualitative comparison of image reconstruction for the various $w$-modulation scenarios considered. In Fig.~\ref{fig:imM31} and Fig.~\ref{fig:im30dor} we show images reconstructed using Daubechies 8 wavelets as sparsity dictionary for different $w$-modulation scenarios. The first row of images corresponds to no $w$-modulation, the second to the varying $w$-component case and the third row shows the reconstructed images with a maximal, constant $w$, \ie\ $w_{\rm d}=1$. Images in the first column are reconstructed images, while those in the second row are the difference of the reconstructed and original image. The residual images in the third column are given by the difference of the dirty images of the original and those of the reconstruction, calculated via $\rm{\bmath x_{\rm{res}}}=\boldsymbol \Phi^{\prime T}\boldsymbol \Phi^\prime \bmath x_{\rm{rec}} - \boldsymbol \Phi^{\prime T}\boldsymbol \Phi^\prime \bmath x$. A relatively low visibility coverage of 10\pc\ is chosen to illustrate differences between reconstruction scenarios clearly.

In Fig.~\ref{fig:imM31} it can be seen that for M31 the reconstructed image without $w$-modulation suffers from considerable contamination, while the reconstructed image for maximal $w$-modulation is relatively free of contamination.  Interestingly, the reconstruction for the more realistic scenario of varying $w$-component (second row) achieves almost the same reconstruction quality as maximal, constant $w$-component (third row).  In Fig.~\ref{fig:im30dor}, the same recovery behaviour as in Fig.~\ref{fig:imM31} can be seen for 30Dor. Again,  the reconstruction for the more realistic scenario of varying $w$-component achieves almost the same reconstruction quality as maximal $w$-modulation with constant $w$.

\begin{figure*}
\subfigure[Dirac basis with M31]{         
                  \centering  
                \includegraphics[width=.45\textwidth]{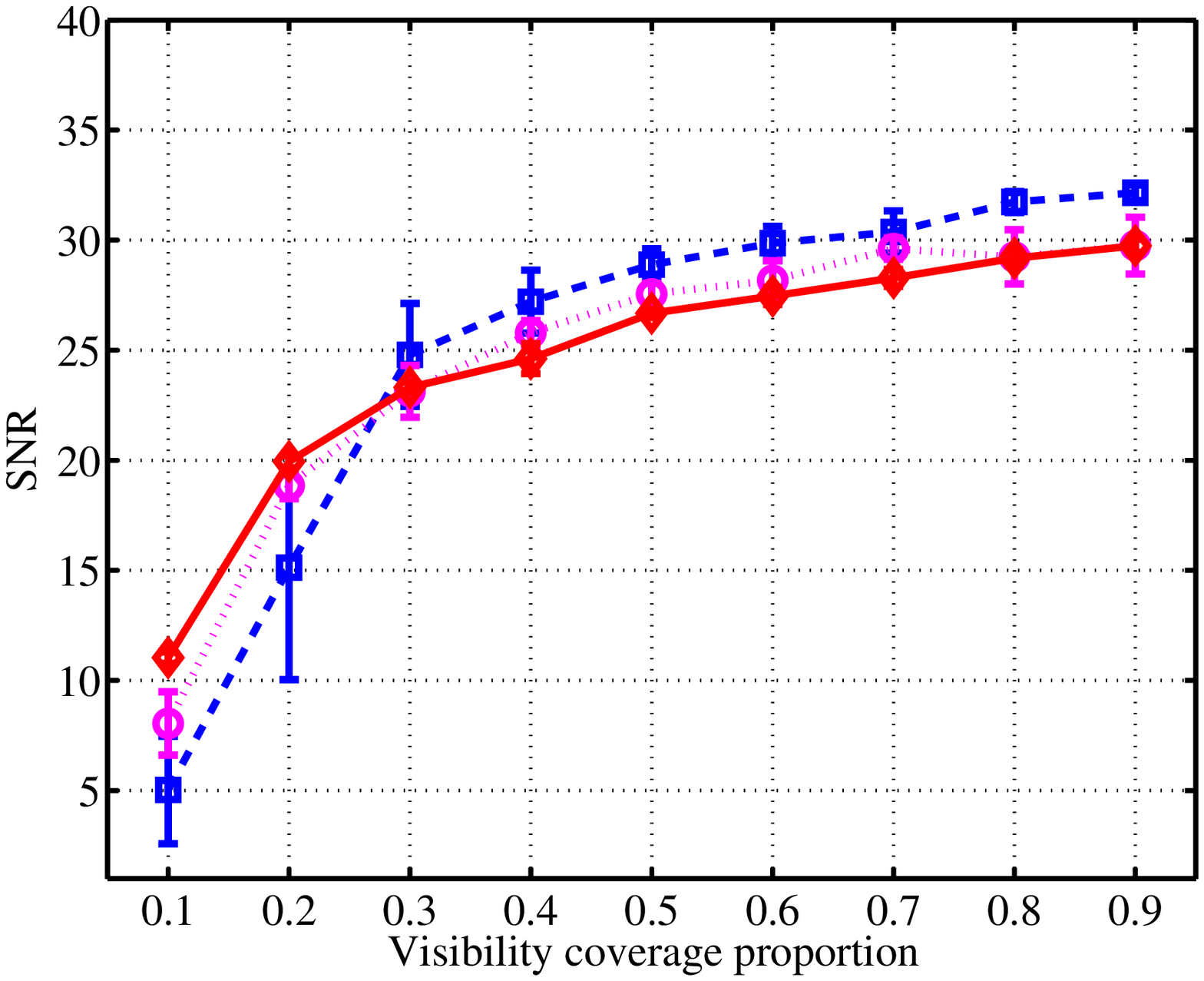}
                \label{fig:diracM31}
        }%
          \subfigure[Dirac basis with 30Dor]{
                \centering
                \includegraphics[width=.45\textwidth]{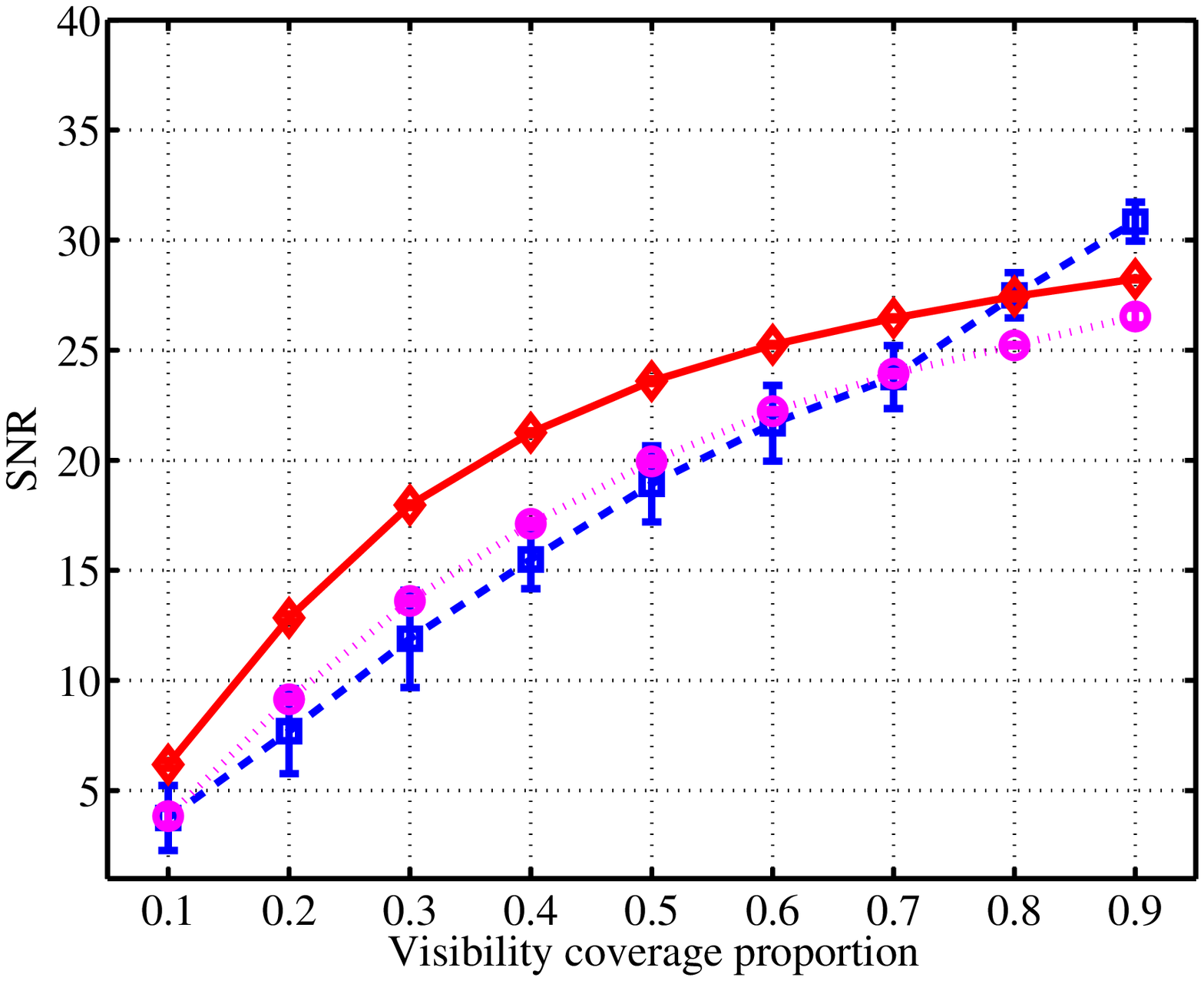}
                \label{fig:dirac30dor}
        }%

\subfigure[Daubechies 8 wavelets with M31]{
                \centering
                \includegraphics[width=.45\textwidth]{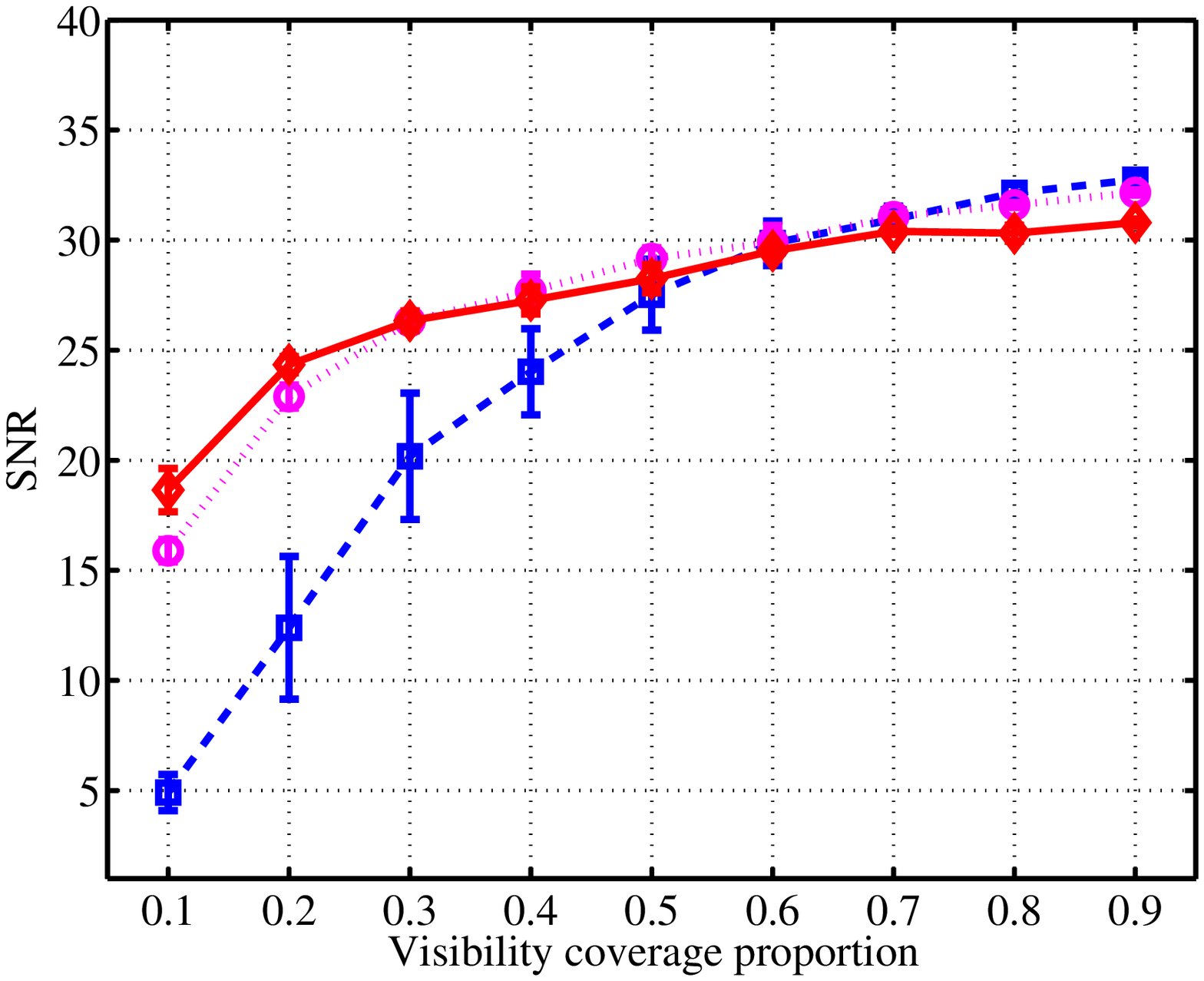}
                \label{fig:db8M31}
	}%
\subfigure[Daubechies 8 wavelets with 30Dor]{
                \centering
                \includegraphics[width=.45\textwidth]{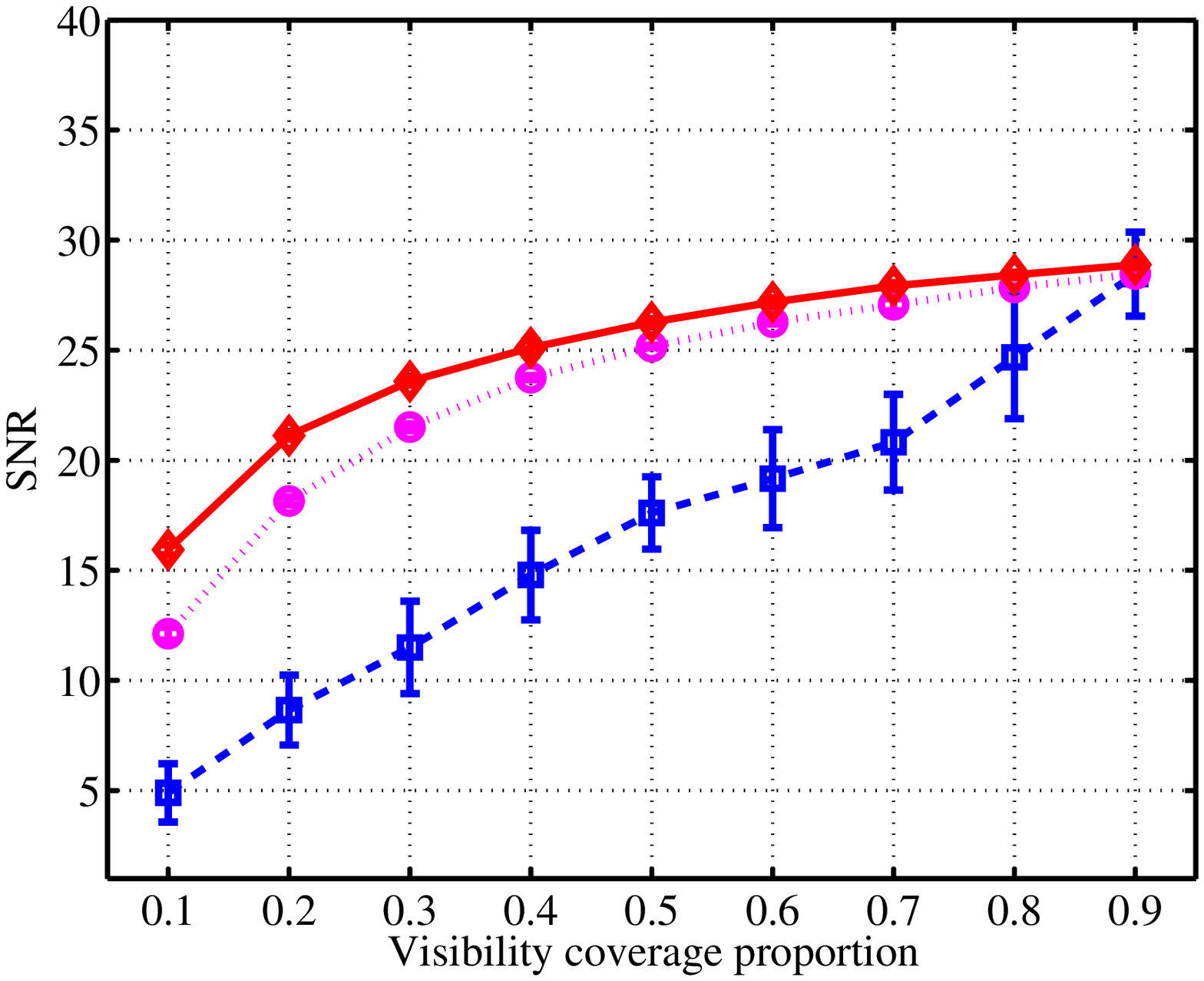}
                \label{fig:db830dor}
        }%

\subfigure[SARA with M31]{
                \centering
                \includegraphics[width=.45\textwidth]{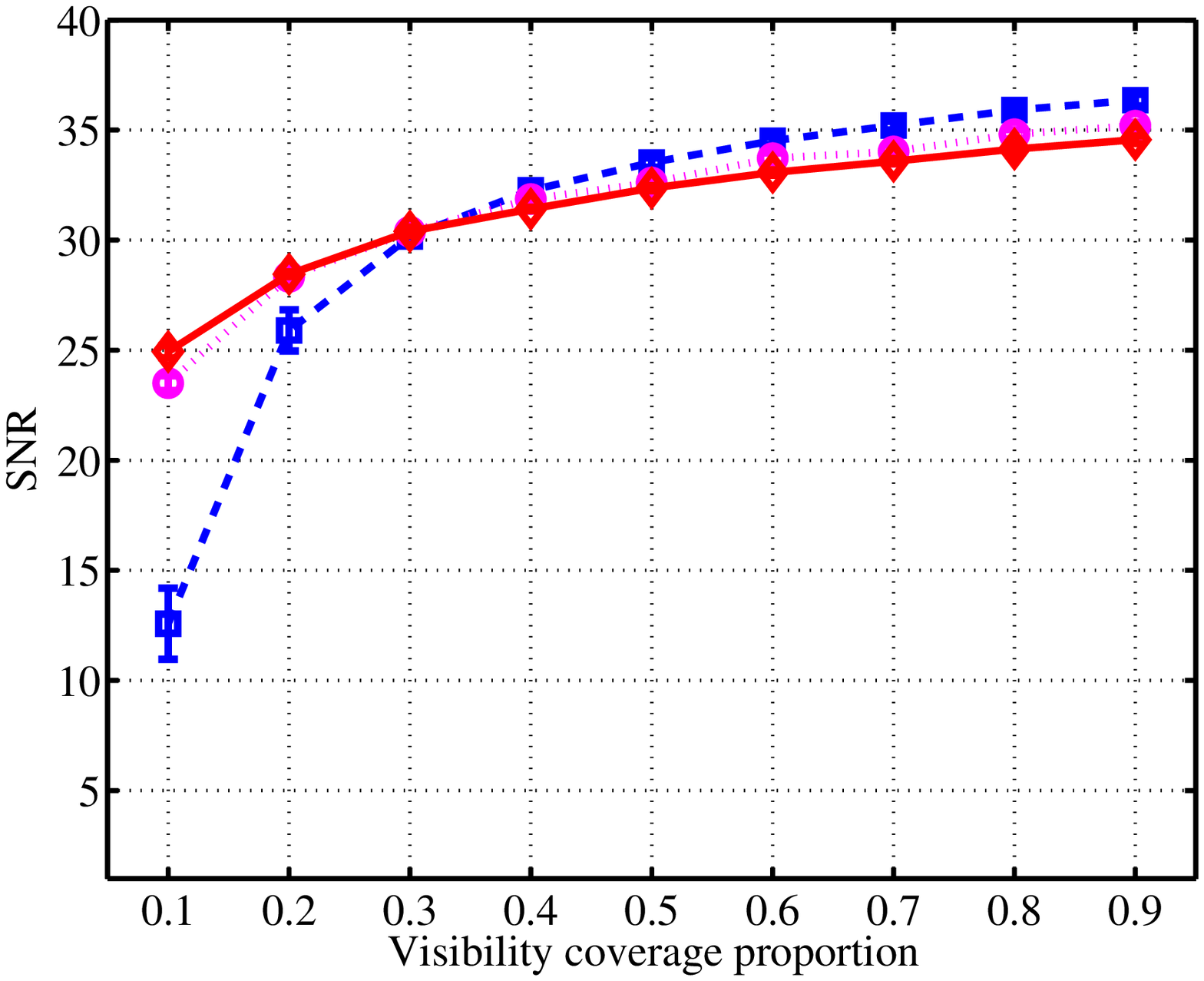}
                \label{fig:saraM31}
        }%
\subfigure[SARA with 30Dor]{
                \centering
                \includegraphics[width=.45\textwidth]{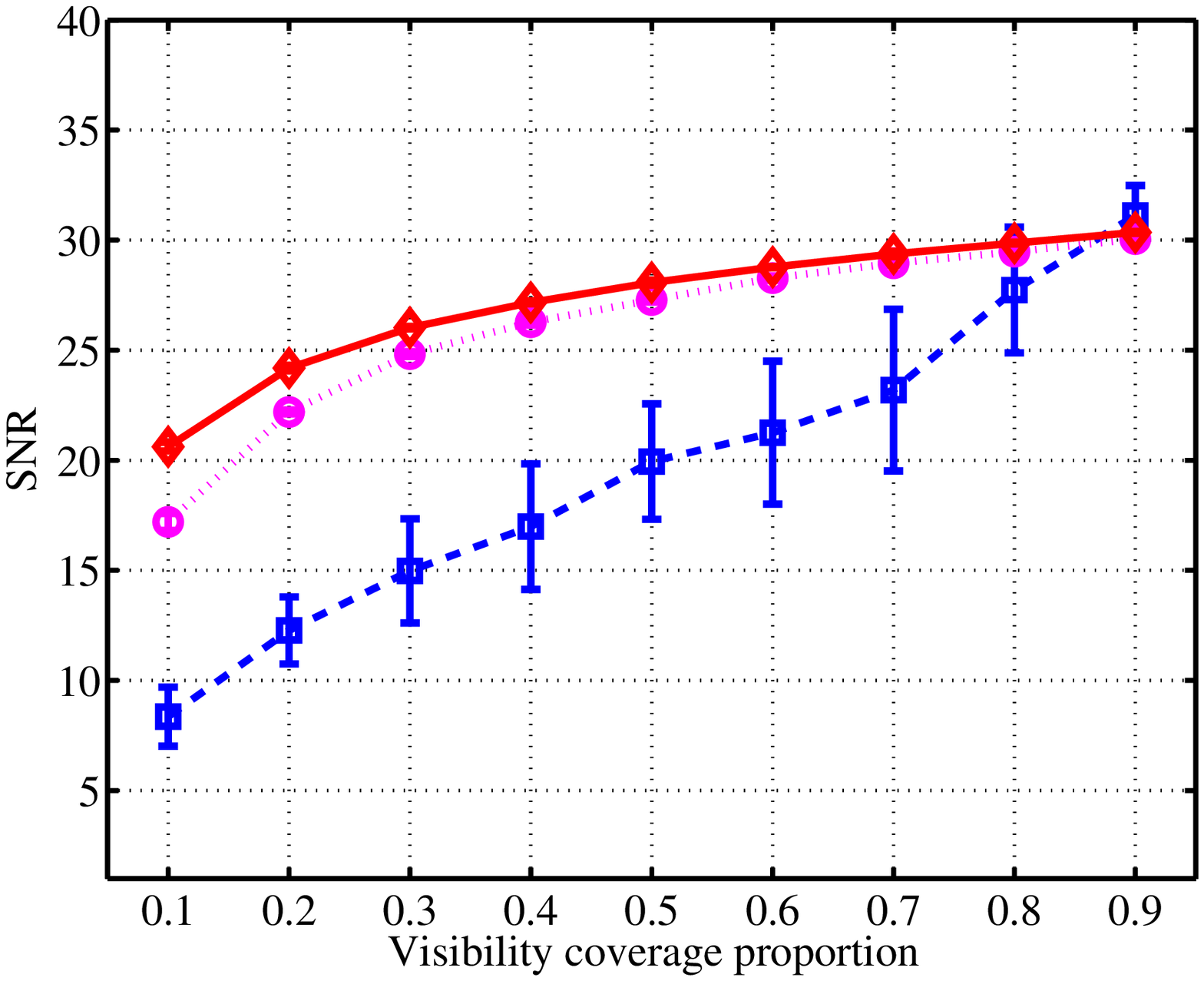}
                \label{fig:sara30dor}
        }
          \caption{Signal-to-noise ratios of the recovered image for no $w$-modulation (blue, dashed lines marked with squares), a varying $w$-component (magenta, dotted lines marked with circles) and a maximal, constant $w$-component (red, solid lines marked with diamonds) as a function of visibility coverage. The input noise level is 30 dB and the energy threshold is chosen as $E=0.75$. The data points are the mean of 10 simulations and the error bars show one standard error. 
        For the Dirac basis in the first row (which is maximally incoherent with the Fourier basis) reconstruction quality remains broadly similar for the various $w$-modulation scenarios, as expected. The quality of reconstructed images using Daubechies 8 wavelets (second row) and SARA (third row) with constant and varying $w$-modulation is higher than that without $w$-modulation.    
           The varying $w$-component performance in these cases is very close to the maximal, constant $w$-component. }     
         \label{fig2}
\end{figure*}

Secondly, we provide a more complete quantitative comparison of reconstruction quality for the different $w$-modulation scenarios, where in addition we consider different sparsity dictionaries and visibility coverages, and average over 10 simulations. In Fig.~\ref{fig2}, the $\rm{SNR_{out}}$ of the reconstructed images are shown as a function of visibility coverage for the two test images. 

When using the Dirac basis as the sparsity dictionary, we do not expect any significant difference in reconstruction quality between the various $w$-modulation scenarios since the measurement basis (essentially Fourier) and the sparsity basis (Dirac) are maximally incoherent. Hence, $w$-modulation will not further improve the recovery of the image. In Fig.~\ref{fig:diracM31} and Fig.~\ref{fig:dirac30dor}, this intuitive understanding is largely confirmed.  In the case of M31, shown in Fig.~\ref{fig:diracM31}, the curves for three $w$-modulation cases are close together and there is no clear tendency that one case performs better than the other over the entire visibility coverage range. In the case of 30Dor, shown in Fig.~\ref{fig:dirac30dor}, the case of a maximal, constant $w$-component seems to perform marginally better than the other two cases; nevertheless, reconstruction quality remains broadly similar.

For Daubechies 8 wavelets and the SARA algorithm on the other hand, we expect an improvement in reconstruction quality due to the $w$-modulation since the sparsity dictionaries are no longer maximally incoherent with the Fourier basis. When using Daubechies 8 wavelets as the sparsity basis, reconstruction quality is improved in the presence of $w$-modulation, as apparent in Fig.~\ref{fig:db8M31} and Fig.~\ref{fig:db830dor}.  Moreover, the curve for the varying $w$-component case is very similar to the maximal, constant $w$-component case. 
For M31 in Fig.~\ref{fig:db8M31}, the varying and maximal, constant $w$-component cases exhibit improved reconstruction quality for low visibility coverages. At coverages above $\sim50$\pc, performance approaches a saturation level approximately equal to the input SNR of 30 dB. For 30Dor in Fig.~\ref{fig:db830dor}, the varying and maximal, constant $w$-component cases show superior reconstruction quality to the absence of any $w$-modulation over a larger range of visibility coverages, until saturation at approximately the input SNR is again reached.  Furthermore, the varying $w$-component and the maximal, constant $w$-component scenarios exhibit very similar reconstruction quality. 
When using the SARA algorithm, the same features are observed as with Daubechies 8 wavelets, as shown in Fig.~\ref{fig:saraM31} and Fig.~\ref{fig:sara30dor}.  Again, it is interesting to note that the varying $w$-component and the maximal, constant $w$-component scenarios exhibit very similar reconstruction quality.  The reconstruction quality with SARA is in general improved compared to Daubechies 8 and Dirac sparsity basis, particularly for very low visibility coverages, as demonstrated by \cite{Carrillo:2012uv} previously.

Fig.~\ref{fig2} shows that reconstruction quality is improved considerably by the spread spectrum effect, caused by the varying $w$-component, for cases where the sparsity dictionary is not the Dirac basis. 
For M31 when using Daubechies 8 wavelets (Fig.~\ref{fig:db8M31}), for example, the same SNR achieved by a scenario without $w$-modulation for a visibility coverage of 40\pc\ can be achieved by 20\pc\ coverage when the spread spectrum effect is considered. For 30Dor (Fig.~\ref{fig:db830dor}), the effect can be seen more significantly where 60\pc\ visibility coverage is necessary to accomplish the same SNR as 20\pc\ coverage when including $w$-modulation.  Similar results are seen for the SARA algorithm although saturation about the input SNR is reached quickly for M31.  If the non-coplanar baseline and wide FoV setting is thus modeled accurately, the same image reconstruction quality can be achieved with considerably fewer baselines due to the spread spectrum effect.

\section{Conclusions}
\label{sec-con}
We have revisited the spread spectrum effect for non-coplanar baseline and wide FoV settings in radio interferometric imaging. We confirmed previous studies (but with more realistic images and alternative sparsifying dictionaries), showing that for $w$-modulation with constant $w$-component,  image reconstruction quality in the context of compressed sensing is enhanced compared to the setting where there is no $w$-modulation (\ie\ $w=0$).  Extending this analysis to the more realistic scenario of varying $w$-component is computationally demanding and is the main focus of the current article.  We developed a variant of the $w$-projection algorithm to support different $w$-components for each visibility measurement, which is based on an sparsification procedure that adapts to the support of each kernel.  Our algorithm is generic and thus suitable for all types of direction-dependent effects.  This approach renders a study of the spread spectrum effect in the varying $w$-component setting computationally tractable.  Through extensive numerical experiments, we find that the image reconstruction quality achieved for the varying $w$-component case outperforms the quality achieved in the absence of $w$-modulation, and is nearly as good as in the case of constant, maximal $w$-component. The enhancement is particularly clear for very low visibility coverages of order a few percent.  

If the non-coplanar baseline and wide FoV setting is thus modeled accurately, the same image reconstruction quality can be achieved with considerably fewer baselines due to the spread spectrum effect.  Alternatively, for a given number of baselines, reconstruction quality is improved by the spread spectrum effect.  This confirms that one may seek to optimise future telescope configurations to promote large $w$-components, thus enhancing the spread spectrum effect and consequently the fidelity of image reconstruction.  Similar suggestions to optimise future telescope configurations to promote large $w$-components have also been made by \citet{Carozzi:2012} recently, but from an information theoretic point-of-view.

In a recent separate work, the scenario is considered where visibilities are measured at continuous $(u,v,w)$ components \citep{Carrillo:2013}, as is the case for real interferometric telescopes.  After integrating these approaches it will be possible to apply our methods and to study the impact of the spread spectrum effect on observations made by real radio interferometric telescopes.

 \section*{Acknowledgments}

LW is supported by the IMPACT fund. JDM is supported in part by a Newton International Fellowship from the Royal Society and the British Academy. FBA acknowledges the support of the Royal Society via a University Research Fellowship. REC is supported by the Swiss National Science Foundation (SNSF) under grant 200020-140861.  YW is supported in part by the Center for Biomedical Imaging (CIBM) of the Geneva and Lausanne Universities, EPFL and the Leenaards and Louis-Jeantet foundations.
\bibliographystyle{mymnras_eprint}
\bibliography{bib}

\label{lastpage}
\end{document}